\documentclass[12pt]{spieman}  
\usepackage{amsmath,amsfonts,amssymb}
\usepackage{graphicx}
\usepackage{setspace}
\usepackage{tocloft}

\title{Review: Machine learning for advancing low-temperature plasma modeling and simulation}

\author[a,b,*]{Jan Trieschmann}
\author[a,c]{Luca Vialetto}
\author[a,d]{Tobias Gergs}
\affil[a]{Theoretical Electrical Engineering, Department of Electrical and Information Engineering, Kiel University, Kaiserstraße 2, 24143 Kiel, Germany}
\affil[b]{Kiel Nano, Surface and Interface Science KiNSIS, Kiel University, Christian-Albrechts-Platz 4, 24118 Kiel, Germany}
\affil[c]{Department of Aeronautics and Astronautics, Stanford University, 496 Lomita Mall, Stanford, CA 94305, United States of America}
\affil[d]{Chair of Applied Electrodynamics and Plasma Technology, Department of Electrical Engineering and Information Science, Ruhr University Bochum, 44780 Bochum, Germany}

\cftpagenumbersoff{figure}
\cftpagenumbersoff{table} 
\begin{document} 
\maketitle

\begin{abstract}
Machine learning has had an enormous impact in many scientific disciplines. Also in the field of low-temperature plasma modeling and simulation it has attracted significant interest within the past years. Whereas its application should be carefully assessed in general, many aspects of plasma modeling and simulation have benefited substantially from recent developments within the field of machine learning and data-driven modeling. In this survey, we approach two main objectives: \emph{(a)} We review the state-of-the-art, focusing on approaches to low-temperature plasma modeling and simulation. By dividing our survey into plasma physics, plasma chemistry, plasma-surface interactions, and plasma process control, we aim to extensively discuss relevant examples from literature. \emph{(b)} We provide a perspective of potential advances to plasma science and technology. We specifically elaborate on advances possibly enabled by adaptation from other scientific disciplines. We argue that not only the known unknowns, but also unknown unknowns may be discovered due to the inherent propensity of data-driven methods to spotlight hidden patterns in data.
\end{abstract}


{\noindent \footnotesize\textbf{*}Jan Trieschmann,  \linkable{jt@tf.uni-kiel.de} }

\begin{spacing}{2}   

\section{Introduction}
Low-temperature plasmas (LTPs) consist of a mixture of neutral and charged species in thermal non-equilibrium. The governing discharge physics and chemistry lead to a zoo of (excited) species interacting with bounding surfaces. Their modification (e.g., activation, etching, functionalization) is exploited in the frame of plasma processing (e.g., plasma-enhanced etching, deposition, catalysis). LTPs also render a dominant part of the many steps required in semiconductor device manufacturing. This may involve a diverse portfolio of plasma discharges, chemical precursors, surface materials, and process recipe steps to achieve desired deposit/etch patterns.\cite{gottschoMicroscopicUniformityPlasma1992, oehrleinFoundationsLowtemperaturePlasma2018} It may as well include \emph{indirect} plasma discharges for example in the generation of extreme ultraviolet light for state of the art 13.5\,nm wavelength lithography, required for device feature scales of the order of 10\,nm and beyond (for instance termed 3\,nm nodes).\cite{tomieTinLaserproducedPlasma2012} Modeling and numerical simulation contribute an integral component in the design and optimization of related plasma processes. In the recent past decade, data-driven methods such as machine learning (ML), deep learning (DL), and artificial neural networks (ANNs), among other computational statistics methods, have (re-)gained enormous interest -- which may be subsumed under the umbrella term artificial intelligence (AI). Although these fields have seen a renaissance and have continued to rapidly develop in the past decade, many fundamental concepts have been well established in computational statistics and related disciplines. The past and present interest in the context of plasma processing may be attributed to the potential for hidden pattern detection in correlated data and, particularly, the efficiency in related optimization tasks.\cite{bishopNeuralNetworksPattern1996, rasmussenGaussianProcessesMachine2005, haykinNeuralNetworksLearning2008, goodfellowDeepLearning2016, geronHandsOnMachineLearning2022, adamovich2022PlasmaRoadmap2022, anirudh2022ReviewDataDriven2023, bonzaniniFoundationsMachineLearning2023, kambaraSciencebasedDatadrivenDevelopments2023}

Although we limit our scope to modeling and simulation in the following, it should be stressed that ML methods and ANNs have been widely employed in experimental studies (e.g., process regression, plasma diagnostics). For instance, as early as in 1992, they have been explored for parameter estimation in plasma etching based on optical emission spectroscopy (OES) and mass spectroscopy, \cite{shadmehrPrincipalComponentAnalysis1992, sangjeenhongNeuralNetworkModeling2003} or global process parameters paired with measured etch characteristics.\cite{himmelAdvantagesPlasmaEtch1993} Plasma virtual metrology (VM) was conducted based on multivariate sensor data,\cite{bleakieGrowingStructureMultiple2016} with regard to real-time fault detection in reactive ion etching,\cite{sangjeenhongNeuralnetworkbasedSensorFusion2005} batch process characterization in semiconductor fabrication,\cite{sutharNextgenerationVirtualMetrology2019} and a deep learning VM framework based on OES data \cite{maggipintoDeepVMDeepLearningbased2019} and 'plasma information' descriptors.\cite{parkMicrorangeUniformityControl2021} Further examples have devised an inverse reconstruction of intrinsic plasma properties such as the electron energy distribution function from OES diagnostics data \cite{vandergaagArbitraryEEDFDetermination2021} as well as an active learning guided scheme based on Fourier transform infrared spectroscopy data for parameter space exploration.\cite{shaoActiveLearningguidedExploration2022}

In contrast to these previous examples which rely on experimental and diagnostics data, the focus of this paper will be on data-driven aspects of plasma modeling and simulation. We consider how these methods can complement and/or replace theoretical approaches, including low and atmospheric pressure LTPs. Specifically, in section\,\ref{sec:review}, we initially attempt to provide a state-of-the-art review of data-driven LTP modeling. This specifically covers plasma physics, plasma chemistry, plasma-surface interactions (PSIs), as well as process control and design aspects. In section\,\ref{sec:advances}, we further relate this to the closely linked disciplines of plasma fusion research, and to more general connections in data science and fluid dynamics research. In this context, an assessment and perspective of future developments and opportunities in LTP research is devised. Overall, the main focus is on DL methods, but also computational statistics methods like Gaussian process regression (GPR) are outlined. Finally in section\,\ref{sec:conclusion}, concluding remarks with a perspective of the current state of the art in adjacent disciplines and the foreseen requirements, potential applications, and future developments of ML in plasma modeling and simulation in the next years 5--10 years is drawn.

\section{Review}
\label{sec:review}
A general perspective on the current state of data-driven LTP science -- among many other aspects -- has been laid out in \emph{The 2022 Plasma Roadmap}\cite{adamovich2022PlasmaRoadmap2022}. Despite the ample momentum that data science and ML have recently gained in LTP, the potential of a community effort for systematic data harvesting and exploration to initiate a form of 'plasma informatics' has been identified therein. Kambara et al.\cite{kambaraSciencebasedDatadrivenDevelopments2023} have further detailed this paradigm with a comprehensive summary of science-based, data-driven developments. A wide range of plasma processing technologies have been addressed, whereas a number of exemplary use cases of data-driven plasma modeling and simulation will be revisited later. Moreover, in an overarching effort the broad scope of the field of data-driven plasma science has been reviewed.\cite{anirudh2022ReviewDataDriven2023} Among other aspects, the fundamentals of data science and its application to space and astronomical plasmas, fusion plasmas, and LTPs have been elaborated. Owing to its broad review scope, LTP processing and its applications have only been generally assessed. In contrast, Bonzanini et al.\cite{bonzaniniFoundationsMachineLearning2023} elaborate in more detail on the fundamentals of ML for LTPs. Therein, the authors not only provide a comprehensive introduction into the concepts and a collection of ML methods with relevant links to the literature, but also underpin their theoretical foundations with an instructive set of applied LTP case studies.

In an attempt to bridge the gap between the mentioned fundamentals of ML -- in general and specifically in the context of LTPs -- and a rather broad overview level, this work focuses on the large variety of physical and chemical aspects of LTP processing that have so far been addressed by data-driven methods (although probably not complete). While we target the topic of modeling and simulation, we specifically consider \emph{(i)} how data-driven plasma science can complement conventional theoretical approaches, as well as \emph{(ii)} when data-driven plasma science approaches may replace classical concepts in the field of LTP modeling and simulation. Further, the interested reader is referred to fundamental literature for a general introduction to data science and ML.\cite{bishopNeuralNetworksPattern1996,rasmussenGaussianProcessesMachine2005,haykinNeuralNetworksLearning2008,goodfellowDeepLearning2016,geronHandsOnMachineLearning2022}

In the following, a review on data-driven methods for plasma modeling and simulation is presented, divided into the major physicochemical characteristics. In subsection\,\ref{ssec:discharge1} approaches to plasma discharge physics phenomena are addressed. Related aspects on plasma chemistry follow in subsection\,\ref{ssec:chemistry1}. Subsection\,\ref{ssec:psi1} deals with plasma-surface phenomena on the mesoscopic and atomistic scale. Finally, subsection\,\ref{ssec:control1} focuses on plasma process control and optimization.

\subsection{Plasma physics}
\label{ssec:discharge1}
LTPs for plasma processing entail a diverse zoo of physical and chemical phenomena with complex interactions. Modeling of such systems must, therefore, comprise at least the governing mechanisms and their intrinsic coupling. For example, consider the spatiotemporal dynamics of charged species coupled to electromagnetic fields. Data-driven modeling of LTP discharge characteristics has followed three prevalent routes: \emph{(i)} Data-driven modeling of a component of a physical model, based on data sets from numerical discharge simulations and finally coupled to the overall numerical simulation. \emph{(ii)} Surrogate modeling that pursues to represent the complete discharge dynamics. \emph{(iii)} Direct embedding of the physical discharge model equations to be solved into a physics-informed ML procedure.

\paragraph{i) Data-driven surrogate sub-modeling} Contributions to this topic have prevalently considered separable submodels as part of a superordinate model consisting of several components. Plasma chemistry presents a specific case detailed later in subsection\,\ref{ssec:chemistry1}. PSI is similarly addressed separately in subsection\,\ref{ssec:psi1}. A major aspect to be considered in the following concerns the description of electromagnetic fields, in particular in electrostatics description. In this case, the Poisson equation presents an elliptic partial differential equation (PDE) subject to time-dependent boundary conditions (BCs). This problem stands out because of its computational effort and difficulties in an efficient parallel solution in multiple spatial dimensions (as the information of the whole domain is required at every iteration).
Early work by Zhang et al.\cite{zhangSolvingPoissonEquation2019} on 2D Monte Carlo (MC) simulations of a PN junction (conceptually related to methods applied in LTP) has considered reference data from the solution of the inhomogeneous Poisson equation subject to Dirichlet BCs using a finite-difference (FD) method. A multi-scale ANN was trained using the internal and boundary values of the electric potential as ground truth. Good agreement paired with a more than 10 times speedup has been reported.
In an attempt to transfer a similar concept to LTPs, Trieschmann et al.\cite{trieschmannMachineLearningApproach2020a} have implemented a 1D self-supervised learning scheme using an FD discretization to establish a physics-based loss, as for instance proposed by Dissanayake et al.\cite{dissanayakeNeuralnetworkbasedApproximationsSolving1994} The data-driven solver was further assessed within a particle-in-cell (PIC) simulation of two collisionless counterpropagating electron streams. A relative error of the order of $10^{-4}$ was reported for the electric potential. However, as the force acting on each particle requires the electric field, evaluation of the first derivative implied significant distortion of the long-term spatiotemporal dynamics. This finding has been corroborated by Aguilar et al.\cite{aguilarDeepLearningBasedParticleinCell2021} who have reported similar findings using fully connected ANNs (a.k.a. multilayer perceptrons (MLPs)\cite{rumelhartLearningInternalRepresentations1985}) and convolutional neural networks (CNNs)\cite{fukushimaNeocognitronSelforganizingNeural1980}.
A 2D solution to Poisson's equation coupled to a particle balance equation for a double headed streamer simulation (in drift-diffusion approximation for electrons and otherwise no transport) has been pursued by Cheng et al.\cite{chengUsingNeuralNetworks2021} (closely related to the work Özbay et al.\cite{ozbayPoissonCNNConvolutional2021}) Therein, a comparison of different ANN architectures was presented, specifically a multi-scale network and a U-Net configuration.\cite{ronnebergerUNetConvolutionalNetworks2015} Whereas a significant speedup has been consistently reported, the accuracy of the predictions was observed to agree only for short term prediction ($<2$ ns). Note that here a simple rectangular domain was considered with the addition of a standardization scheme to extend to variable domain sizes. Further, the loss was composed of the inner domain and Dirichlet boundary values, paired with a physics-based loss representing the Laplacian.
Zeng et al.\cite{zengPhysicsInformedNeural2023} have followed a different paradigm using physics-informed neural networks (PINNs)\cite{raissiPhysicsinformedNeuralNetworks2019,luDeepXDEDeepLearning2020}. Therein, physical coordinates are input to an ANN and the underlying PDE is implemented as a physics-informed loss function. Generally, this loss function may include a set of possibly coupled PDEs to be solved within the domain, the BCs, and the initial conditions in case of transient phenomena.\cite{raissiPhysicsinformedNeuralNetworks2019,luDeepXDEDeepLearning2020} Automatic differentiation is exploited to obtain the respective local error gradients with respect to the input variables.\cite{baydinAutomaticDifferentiationMachine2018} It resembles a paradigm closely linked to the backpropagation algorithm\cite{rumelhartLearningInternalRepresentations1985} for training ANNs and enables a mesh-less representation of the PDE solution. In the referenced work, the inhomogeneous Poisson equation with rectangular and circular dielectric objects was solved in two and three dimensions, subject to time-dependent Dirichlet and/or Neumann BCs. Note that the authors have initially included time as a variable to consider time-dependent BCs (but later reverted to a time-invariant formulation), although Poisson's equation essentially poses a static PDE problem. Hence, the error metric was composed of a domain loss representing the Laplacian and the loss corresponding to Dirichlet BCs. Whereas the application to dielectric barrier discharges (DBDs) was proposed as use case, an integration into a complete LTP model remains due.
An approach dedicated to the requirements of complex LTP simulations has been put forward by Siffa et al.\cite{siffaMachineLearnedPoissonSolver2023} based on a combination of U-Net,\cite{ronnebergerUNetConvolutionalNetworks2015} DenseNet,\cite{huangDenselyConnectedConvolutional2017} and transformer\cite{vaswaniAttentionAllYou2017,chenTransUNetTransformersMake2021} network architectures, called TransDenseUNet. On a 2D Cartesian mesh, the non-overlapping domain information was encoded into types (plasma, dielectric, BCs), corresponding physical values (dielectric constant, charge density, boundary potential/normal derivative), and spatial coordinates. A feature scaling of the physical equation was proposed and all input/output quantities were normalized prior training. The data set was composed of a few physical reference cases for intended applications and a substantial amount (10k to 50k) of randomly generated data samples. The training loss was combined with various contributions (including a gradient loss and a structural similarity index measure (SSIM)\cite{wangImageQualityAssessment2004}) to increase the accuracy and the smoothness of the electric potential prediction (and the associated electric field). The proposed data-driven model was demonstrated to successfully generalize on arbitrary geometries, BCs, and charge distributions -- including realistic cases encountered in LTP processing. Also here, an implementation within a complete plasma model remains due.

With the goal of studying the complex interaction of electromagnetic waves with plasma, Desai et al.\cite{mihirdesaiDeepLearningArchitectureBasedApproach2022} have suggested a U-Net architecture to predict the 2D distribution of the root mean square electric field based on plasma density distributions. The ground truth data set was obtained from finite-difference time-domain (FDTD) simulations. Interestingly, while a simple regularized L2 loss was used, the quality of the prediction was assessed using the SSIM and the average percentage error. Whereas overall good agreement with reference simulations was achieved, an issue generally noticeable with many ANN surrogate models was observed. Specifically, the convergence toward a reliable prediction is subject to a spectral bias. Lower frequency features converge significantly faster than higher frequency features. Consequently, it may be computationally intractable to train until convergence is achieved for all relevant frequencies.\cite{rahamanSpectralBiasNeural2019}

\paragraph{ii) Data-driven discharge surrogate modeling} An early example by Verma et al.\cite{vermaSurrogateModelsLow2021} has considered data sets from 1D time-resolved fluid simulations of capacitively coupled radio-frequency discharges to establish surrogate models for integrated quantities such as the plasma sheath voltages/currents, that capture reduced temporal modes of operation. In contrast to physical model based solutions, the authors have highlighted the equation-free representation of their long short-term memory (LSTM, cf. subsection\,\ref{ssec:concepts2})) surrogate model. The corresponding reduced order description and its latent space representation was suggested for parameter space exploration. This may otherwise be infeasible due to computational requirements.
Another approach to capacitively coupled plasma surrogate modeling by Hamaguchi and co-workers has been devised based on 1D fluid and particle-in-cell with Monte Carlo collisions (PIC/MCC) simulations of an Ar discharge.\cite{ichikawaConstructionSurrogateModel2022} Instead of a grid-based approach for training and prediction (where the whole spatial domain is regressed on at once), MLPs were trained on phase-average plasma quantities sampled at varying positions within the discharge. The dynamics of an inductively coupled plasma with additional radio frequency substrate bias has been implemented based on HPEM (hybrid plasma equipment model)\cite{kushnerHybridModellingLow2009} simulations by Ko et al.\cite{koComputationalApproachPlasma2023} 2D axisymmetric simulations of an Ar etch plasma were performed with a fixed geometry (rectangular mesh) and varying process conditions (pressure, source power, bias power at 1\,MHz and 13.56\,MHz). The devised data set was further used to train a multi-encoder/decoder ANN, which encodes the process variables and geometry into a latent representation to finally decode the corresponding phase-averaged discharge quantities. Note that although the geometry was fixed in the presented study, the proposed ANN architecture with dedicated geometry inputs (sources, material and properties) may also be capable of taking varying geometrical features (e.g., chamber aspect ratio, wafer diameter) into account. Due to its low computational costs, the ML surrogate model allowed for a rapid exploration and optimization of a uniform and sufficient etch rate using the multi-objective particle swarm optimization algorithm.

Data-driven surrogate modeling applied to different atmospheric discharges has been proposed by Zhang et al.\cite{zhangEfficientNumericalSimulation2023a,zhangEfficientNumericalSimulation2023,wangModelingDischargeCharacteristics2023} A drift-diffusion model taking into account particle balance coupled to Poisson's equation, and the electron energy conservation equation was used to simulate pulsed He and CO$_2$ discharges. Corresponding data sets were established for varying source properties and subsequently used for setting up an MLP. A reduced set of input descriptors was selected, such as time and source properties (e.g., voltage, pulse rise rate). As output descriptors the discharge current density or the spatial distributions of species were chosen. With reference to these simulations (and experiments) it was demonstrated that a rather simplistic ANN even in the absence of complex learning schemes is capable of capturing the spatiotemporal discharge dynamics as well as the chemical reaction kinetics. 

\paragraph{iii) Physics-informed machine learning} PINNs were further used to directly consider the equations governing the physics of LTP discharges. Similar to the previously mentioned PINN solution to Poisson's equation, corresponding PDEs may be implemented in the error metric during training.\cite{raissiPhysicsinformedNeuralNetworks2019,luDeepXDEDeepLearning2020} Automatic differentiation is again exploited to obtain desired partial derivatives with respect to the input variables.\cite{baydinAutomaticDifferentiationMachine2018}
Early work in LTP has been proposed by Kawaguchi et al.\cite{kawaguchiDeepLearningSolving2020} who have considered a PINN solution to the stationary Boltzmann equation subject to ionization and electron attachment processes. An MLP has been trained to predict the electron velocity distribution function (EVDF) following an exponential ansatz. The solution was validated for two reference gas mixtures, whereas the dependence on a variation of the reduced electric field was used as a case study. Excellent agreement between the data-driven approach and MC simulations was reported. Several improvements to their initial scheme have been suggested by the authors. Specifically, a nondimensionalization of the equation combined with an improved ANN architecture was used.\cite{kawaguchiPhysicsinformedNeuralNetworks2022} The last is a fully connected ANN which includes two transformer layers for an improved training of the PINN.\cite{wangUnderstandingMitigatingGradient2021,vaswaniAttentionAllYou2017} Further, a scheme was proposed to take into account variable reduced electric field values $E/N$.\cite{kimNumericalStrategySolving2023} In both studies, overall good agreement with MC simulations was demonstrated. For Ar gas eight processes were included: momentum transfer, six electronic excitations and ionization. For SF$_6$ gas nine processes were included: momentum transfer, vibrational and electronic excitations, electron attachment and ionization. An extension to the time-dependent Boltzmann equation and assessment of the memory capacity of the ANN of 0.1\% compared to mesh-based DNS calculations of EVDFs in 3D velocity space was further argued.

In the frame of a PINNs for LTP, Zhong et al.\cite{zhongDeepLearningThermal2020} have initially investigated stationary and transient 1D arc discharges. A fully connected ANN was used to implement the stationary Elenbaas--Heller equation (with decay term) as well as the time-dependent Elenbaas--Heller equation (considering mass and energy conservation). Whereas the PDE was generally solved using PINNs, the thermodynamic, transport, and radiation properties of the plasma were (partially) taken into account via so-called coefficient subnets. These resemble small regression networks to interpolate the corresponding plasma properties. In a follow-up study,\cite{zhongLowtemperaturePlasmaSimulation2022} the authors have introduced this framework more generally in the frame of Coefficient-Subnet Physics-Informed Neural Network (CS-PINN) and Runge–Kutta Physics-Informed Neural Network (RK-PINN). For the first approach, a comparison with cubic spline interpolation was included. The last approach embeds PINNs in an implicit Runge–Kutta formalism. In a case study for three examples of Boltzmann equation (with ionization, electron attachment, elastic collision, and excitation), Poisson-drift-diffusion equations (for electrons and ions), and the previously addressed 1D arc discharge using the time-dependent Elenbaas--Heller equation, excellent agreement was noted. Zhong et al.\cite{zhongAcceleratingPhysicsinformedNeural2023} have lastly revisited the data-driven 1D arc discharge simulation, introducing the concept of meta-learning.\cite{psarosMetalearningPINNLoss2022} Therein, a meta loss was aggregated over multiple tasks (e.g., parameters governing discharge conditions, time-dependence) to enable an efficient two-step optimization and, therefore, reduce the training time. For each training task, a regular PINN was trained. In contrast, the meta network was trained to provide good initial weights for a PINN solving a new task. Based on a variation of arc discharge parameters and corresponding meta-training, a substantial acceleration with a factor in the range of 1.1 to 6.9 was reported.

A different approach to PINNs following Raissi et al.\cite{raissiPhysicsinformedNeuralNetworks2019} involves the discovery of PDEs from observations. A single PDE or a set of PDEs may be formulated as prototype equations. Therein, unknown coefficients $\lambda_i$ determine the exact form of the equations. The problem can be phrased such that training a PINN not only yields the desired solution to the PDE, but includes the coefficients $\lambda_i$ as parameters of the PINN. This concept has been used by Kawaguchi et al.\cite{kawaguchiDatadrivenDiscoveryElectron2023} to ``discover'' electron transport coefficients from electron swarm maps measured by scanning drift-tube experiments. The temporal and spatial evolution of the electron swarm was formulated in two PDEs with corresponding transport coefficients. A fully connected ANN with transformer layers was utilized similar to previous works by the authors.\cite{kawaguchiPhysicsinformedNeuralNetworks2022} The tunable parameters are optimized along with the ANN, without assuming any analytical form of the electron swarm map. It was concluded that more accurate transport coefficients could be obtained, including higher-order dependencies if required.

Xiao et al.\cite{xiaoMultiscaleModelingRecurrent2021, xiaoRecurrentNeuralNetworkBasedModel2022} have considered a data-driven surrogate model based on recurrent neural networks (RNNs) and a reduced-order model (ROM) for model predictive control (MPC) of a plasma etch process. It is therefore addressed more specifically in subsection\,\ref{ssec:control1}.

\subsection{Plasma chemistry}
\label{ssec:chemistry1}

Low-temperature plasma models are based on a set of chemical reactions which describes volumetric interactions between all the species tracked in the model, such as electrons, ions, and neutrals.\cite{gravesReportScienceChallenges2023} An accurate description of such interactions is required in order to 
understand and optimize the selective generation of chemically active species and their properties, 
such as mean energy and velocity distribution. 
For plasma modeling applications, chemistry sets may include up to hundred species and several thousands of reactions.\cite{turnerUncertaintySensitivityAnalysis2016} 
Such chemical complexity can hinder the process of understanding and optimizing plasma applications. Furthermore, the plasma modeling community is sometimes facing scarce availability of fundamental data, such as cross sections and reaction rate coefficients, 
that are used as input in the models.\cite{alvesFoundationsPlasmaStandards2023}
With significant advances of ML, new tools are available to the community to improve plasma chemistry models. 
There are three major roles of ML with respect to plasma chemistry developments, that are: 
\emph{(i)} extraction of cross sections and reaction rate coefficients, \emph{(ii)} derivation of reduced reaction mechanisms, and \emph{(iii)} chemical engineering. In particular, the following questions are addressed in the present review: 

\begin{itemize}
\item How can electron-impact cross sections with atoms and molecules of interest be rapidly computed?
\item How can ML enable developments of reduced reaction mechanisms for plasma chemistry? 
\item How to efficiently engineer chemical reactions systems with ML?
\end{itemize}

\paragraph{i) Extraction of cross sections and reaction rate coefficients}

\begin{figure*}[tb!]
    \centering
    \includegraphics[width=0.8\textwidth]{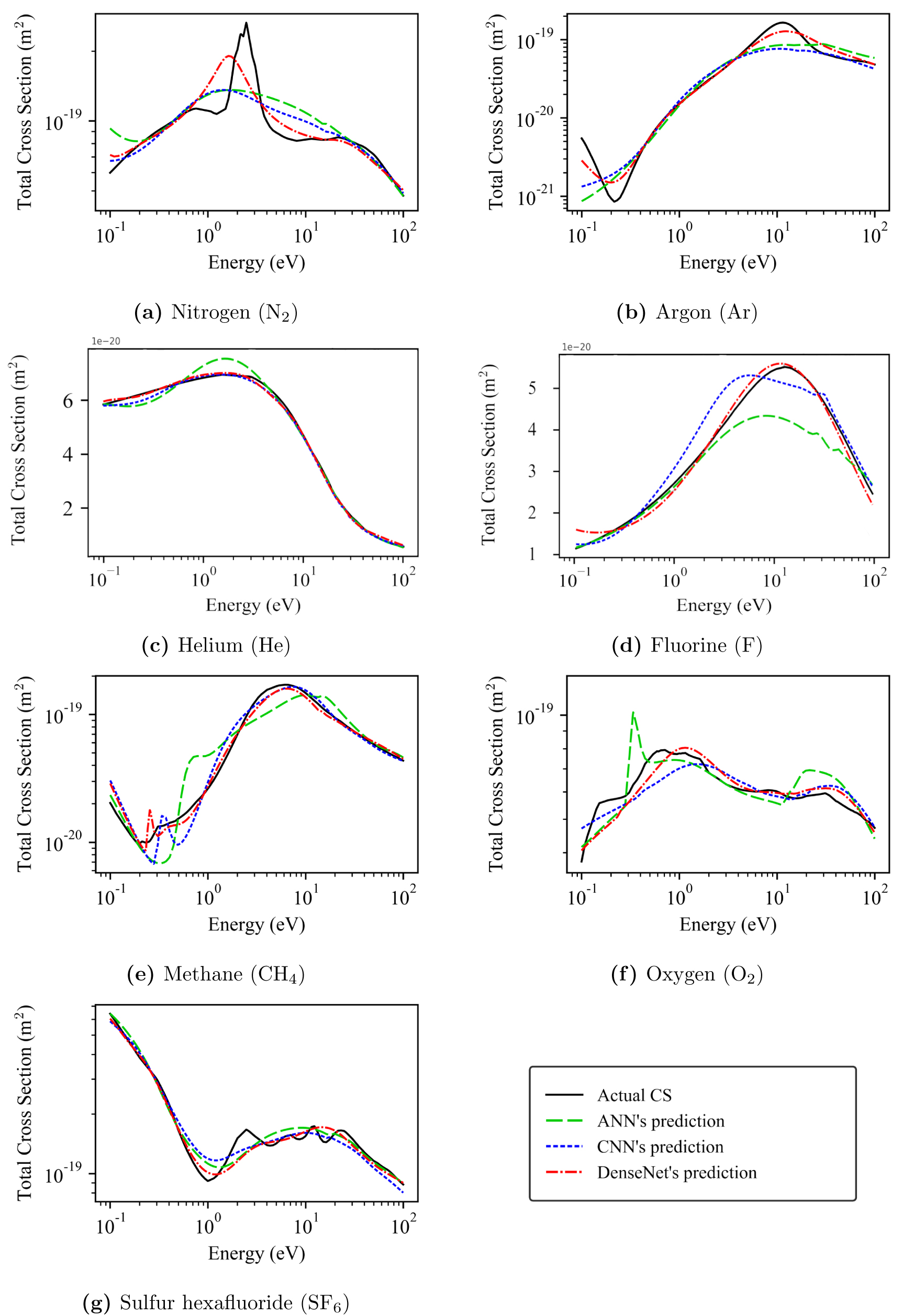}
    \caption{Predicted total momentum transfer cross sections of electrons with various gas species. Figure reproduced from Jetly and Chaudhury\cite{jetlyExtractingElectronScattering2021} licensed under \href{http://creativecommons.org/licenses/by/4.0/}{CC BY 4.0}.}
    \label{fig:cross_sections}
\end{figure*}

The use of an ANN as an optimization technique for treating the inverse problem of obtaining 
electron collision cross section from electron transport coefficients has been proposed by Morgan.\cite{morganUseNumericalOptimization1991} In that pioneering work, it was assessed that neural networks are indeed useful to determine the elastic momentum 
transfer cross sections of electrons in real gases, but the accuracy level is highly dependent on the 
quality of input cross section data that are used for training of the model. 
Stokes and co-authors\cite{stokesDeterminingCrossSections2020, stokesImprovedSetElectronTHFA2021} have improved the idea by Morgan and implemented deep neural network (DNN) using cross sections from the LXCat database\cite{lxcat-teamHttpsNlLxcat2023} associated with electron transport coefficients 
found by numerical solutions of the electron Boltzmann equation or by experimental measurements. 
In their works, the authors demonstrated that their automatic solution using ANN had an accuracy comparable to that of a
human expert in determining cross sections of electrons in He or in the biomolecule tetrahydrofuran (THF). 
The authors have also shown that the use of large amount of synthetic training data generated by using the real cross
sections available from LXCat provide good results for the prediction of elastic momentum transfer and
ionization cross sections of He and Ar.\cite{stokesImprovedSetElectronTHFA2021} 
However, the architecture of the ANN used in ref.\cite{stokesDeterminingCrossSections2020} and ref.\cite{stokesImprovedSetElectronTHFA2021} had minor
improvements over the architecture proposed by Morgan\cite{morganUseNumericalOptimization1991}. 
Jetly and Chaudhury \cite{jetlyExtractingElectronScattering2021} have implemented an ANN, specifically CNN and densely
connected convolutional network (DenseNet)\cite{huangDenselyConnectedConvolutional2017} to derive electron impact cross sections from a solution of the 
inverse electron swarm problem. The networks were trained using elastic
momentum transfer, ionization, and excitation cross sections for different gases available in the
LXCat database\cite{lxcat-teamHttpsNlLxcat2023} and their corresponding electron transport coefficients, which were calculated using the BOLSIG+ solver.\cite{hagelaarSolvingBoltzmannEquation2005} 
For the first time, the performance and accuracy in cross sections determination from different 
network architectures was assessed and it was found that DenseNet predicts cross sections with significantly higher accuracy compared to other ANN models due to its ability to extract both long and short term features from the electron transport parameters. 
Figure\,\ref{fig:cross_sections} shows that the DenseNet model, as opposed to other ANN-based models (e.g., CNN), is able to predict specific cross sections features 
in energy, such as the resonance of the elastic momentum transfer cross sections of electrons in N$_2$. 
Improvements of such deep learning models, such as the ability to extract multiple excitation cross sections and/or the use of sophisticated synthetic data generation schemes, may facilitate the use of ML-derived electron-impact cross sections in plasma chemistry models.

Fast prediction of collision cross sections based on simulated cross sections has been initially suggested by Zhong.\cite{zhongFastPredictionElectronimpact2019} While the author used density functional theory (DFT) calculations to devise a data set for electron impact ionization cross sections for molecules, it has been argued that the approach would be similarly applicable to experimentally measured cross sections. Based on a support vector machine (SVM) model with radial basis function kernel, the prediction of accurate cross sections has been verified. Moreover, the capability to generalize from small to large molecules has been argued, without the need for computationally costly DFT calculations.
Following a similar paradigm, Harris and Nepomuceno have recently developed a conceptually simple but effective ANN model to predict electron-molecule ionization cross sections from a database of measured cross sections.\cite{harrisDataDrivenMachineLearning2023} In particular, by exploiting the chemical configuration (i.e., number of atoms in a molecule) as input to the network, the authors have evaluated the capability of a generalized prediction of cross sections for additional molecules.

Direct determination of reaction rate coefficients and their temperature dependence is also important in plasma modeling. In fact, each reaction needs to be assigned with its own kinetic parameters and these are sometimes not available or subject of large uncertainties. 
Reiser and co-authors \cite{reiserDeterminingChemicalReaction2021} have used ML to derive rate coefficients for radio-frequency atmospheric pressure plasmas operated with various gas mixtures of O$_2$, CH$_4$, and He.
In their work, the authors used a genetic algorithm to derive rate coefficient values that bring the results
of the integrated set of particle balance equations for different species in the plasma into good agreement 
with the experimental data. 
Even if this approach could be prone to biases induced by user experience and be limited by the availability of experimental data, it has been demonstrated that results from the genetic algorithm 
open up the possibility of learning more about the significance and correlation of individual reactions.
Hence, this method is potentially interesting for post-processing and evaluating experimental data using simplified reaction schemes. 
Hanicinec and co-authors \cite{hanicinecRegressionModelPlasma2023} have explored the use of ML to supply unknown
reaction rates in plasma chemistries, thus allow complete
chemistry sets to be generated without resorting to estimations by analogy or an educated guess. 
In their work, ML was used to provide rate coefficients of binary chemical reactions using 
three distinct optimized regression models: An SVM model, 
a random forest model, and a gradient-boosted trees model. 
The models were trained on kinetic data for binary heavy-species collisions
at or near room-temperature extracted from the QBD \cite{tennysonQDBNewDatabase2017}, KIDA \cite{wakelamKineticDatabaseAstrochemistry2012}, NFRI \cite{parkNewVersionPlasma2020}, and UfDA \cite{mcelroyUMISTDatabaseAstrochemistry2013} databases.
After removing duplicate reactions, the final data set consisted of 9470 reactions involving 1080 distinct species.
As a sample use case, the ML results were used to augment the chemistry of a BCl$_3$/H$_2$ gas mixture.
In this example, reactions between the various BCl$_x$ species and H were missing. 
Results of their work showed the importance of taking such reactions into consideration for which the ML algorithm likely gives better
estimates than intuitive guesses, while being faster than measurements 
or ab initio calculations.\cite{hanicinecRegressionModelPlasma2023}

\paragraph{ii) Derivation of reduced reaction mechanisms}

Several methods have been developed to provide interpretation and 
obtain proper reduction of complex chemical kinetic sets. 
There are two main methods for reducing reaction mechanisms: 
reparametrization of the chemical state space and graph theory related algorithms. 
For the first category, we can refer to principle component analysis (PCA) \cite{jolliffePrincipalComponentAnalysis2016} and intrinsic low-dimensional manifold (ILDM) \cite{maasSimplifyingChemicalKinetics1992} as two methodologies that are used for LTP kinetics reduction. 
In PCA, a manifold is generated from linearization of the chemical source terms. 
Reduction is obtained since the chemical state space that is described by a small
number of parameters, the principal components (PCs), and balance equations for the PCs 
are solved instead of the full set of equations. 
Peeremboom and co-authors \cite{peerenboomDimensionReductionNonequilibrium2015} have applied PCA to a 0D state-to-state kinetic
model of CO$_2$ including vibrational levels of CO$_2$ and CO.
The a priori analysis showed that log-transformation and scaling are important for 
the manifold identification. Moreover, a significant speedup in central processing unit (CPU) time has been obtained 
due to the fact that PCA reduced not only the number of variables, but also the stiffness of the 
equations. 
The ILDM method has been applied by Rehman and co-authors \cite{rehmanSimplifyingPlasmaChemistry2016} to Ar and H$_2$ plasma kinetic schemes. 
It has been shown that the ILDM method automatically extracts
the relevant information about fast and slow time scales in a chemical reaction
system. After a short interval of time the fast time scale processes will quickly move towards this low-dimensional manifold and the slow time scale processes will move tangential or along the
manifold to finally reach the equilibrium point.\cite{rehmanSimplifyingPlasmaChemistry2016} 
Despite successful application in plasma chemical systems, PCA and ILDM also presents some 
drawbacks. For example, in PCA, the manifolds are constructed based on the hypotheses 
of linearity between original and transformed variables that cannot be always satisfied for 
complex chemical networks. Moreover, ILDM can be problematic for chemical kinetics that have more
than one equilibrium point or for more complex systems where the increase in the dimensionality of the manifolds and subsequently of the generated look-up tables can require higher computational resources. 

\begin{figure*}[tb!]
    \centering
    \includegraphics[width=\textwidth]{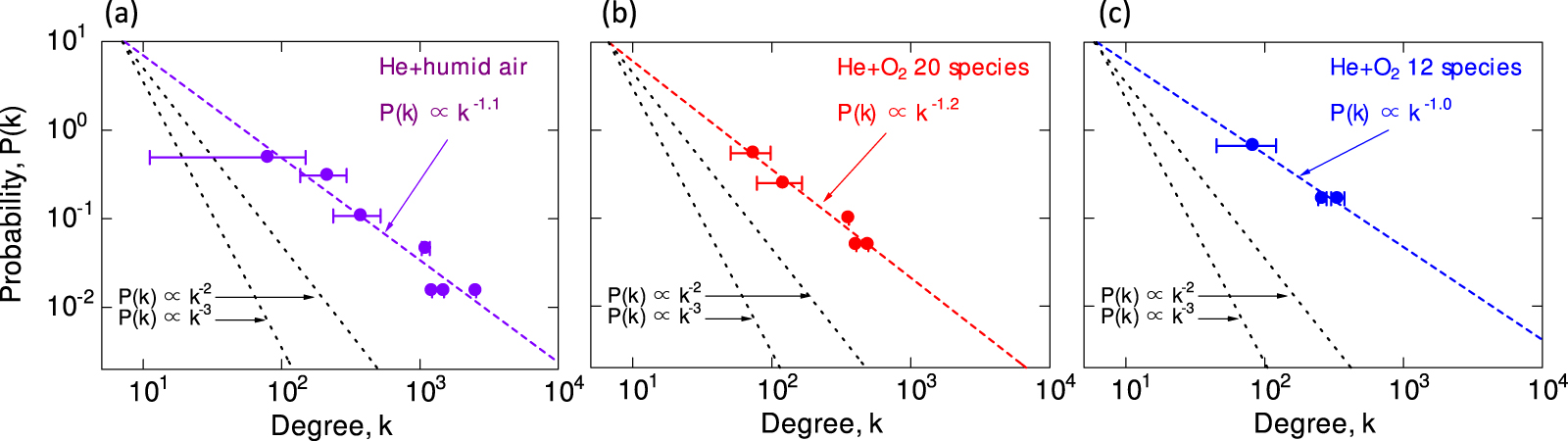}
    \caption{Power-law distributions of the number of degrees for (a) He + humid air plasma, (b) He + O$_2$ plasma, 20-species model and (c) He + O$_2$ plasma, 12-species model. Figure reproduced from Murakami and Sakai\cite{murakamiRescalingComplexNetwork2020} licensed under \href{http://creativecommons.org/licenses/by/4.0/}{CC BY 4.0}.}
    \label{fig:scale_free_network}
\end{figure*}

Graph theory has been applied to chemical kinetics for visualization, topological statistics calculation of a chemical network, and determination of pathways from a sequence of reactions.
Sakai and co-authors\cite{sakaiAnalysisWeblikeNetwork2015} have been the first to apply it to complex plasma chemistry. 
In their work, they have constructed a network structure for methane
and silane plasmas by representing the species as nodes and reactions as directed unweighted edges and 
they demonstrated that this network analysis allows one to estimate substantial effects of given species on other ones existing
in the network.
Then, the work was extended with betweenness and centrality indices to highlight not only the influentiality of each species on/to the others, but also global structures in chemical networks.\cite{sakaiComplexityVisualizationDataset2022} 
In this respect, these works have opened up possibilities for simpler prediction of chemical reactions that one would like to understand and/or control. 
A similar approach for graph construction has been adopted by Sun and co-authors,\cite{sunChemistryReductionComplex2020} with a different choice of undirected edges
weighted based on steady-state reaction rates. The reduction method has been applied to CO$_2$ chemical kinetics for a gliding arc plasma. 
Murakami and Sakai\cite{murakamiRescalingComplexNetwork2020} have used a graph-based approach for extracting a low-dimensional reaction set based on centrality indices. 
In particular, their approach was applied to He$-$O$_2$ and He$-$humid air plasma
chemistry and they have shown that a reduced chemical reaction set for these systems can be achieved with the help of topological centrality (closeness and betweenness) and scale-freeness. In particular, in Figure\,\ref{fig:scale_free_network}, it is shown that a plasma-induced reaction network can be described with a scale-free network, where the number of degrees originating from a particular species follows a power-law distribution. 
The advantage of their approach is that it can be used as preprocessing to reduce complex plasma chemistry before multi-dimensional simulations. 
Mizui and co-authors\cite{mizuiGraphicalClassificationMultiCentralityIndex2017} have extended the analysis in ref.\cite{murakamiRescalingComplexNetwork2020} to transient states by constructing temporal graphs. 
Their classification method was used to deduce information from silane and methane plasma chemistries.
In particular, quantitative insights into chemical mechanisms have been derived with axes of multi-centrality indices.
Holmes and co-authors\cite{holmesGraphTheoryApplied2021} used a direct graph representation of plasma mechanisms 
in which both species and reactions are identiﬁed as nodes, with the edges weighted based on the corresponding reaction
rate coefﬁcients. 
Furthermore, they have used a connectivity matrix to get qualitative information on fast and slow reaction pathways. 
With this approach, they have shown that the new approach can produce a clearer image for the interpretation of complex chemistry. 
Venturi and co-authors\cite{venturiUncertaintyawareStrategyPlasma2023} have presented a novel method for reducing plasma chemistry mechanisms by taking into account the
uncertainties affecting the reaction parameters. As a difference with respect to previous works, their approach relies on an 
ensemble of weighted directed graphs having both species and reactions as nodes. 
Edges are weighted based on the reaction contribution to the particle balance equation.
The methods were applied to a plasma mixture of 23 species including 15 vibrational states of the H$_2$ molecule, three
electronic states of the H atom, four ions, and electrons. As a result of their work, 
it has been shown that a graph ensemble signiﬁcantly improves the methodology’s robustness while preserving the predictive accuracy.
In general, graph theory related algorithms have been successfully applied to find patterns in large chemical data set. These algorithms could be used to identify and analyze patterns in chemical kinetic data prior running experiments or simulations \cite{venturiUncertaintyawareStrategyPlasma2023}.

\paragraph{iii) Chemical engineering}

In many applications, the plasma community is interested in 
knowing ``how we can'' generate a specific concentration of species. 
In this respect, modeling and simulations are powerful tools to allow one to actively 
modulate plasmas and/or discharges for production of chemically reactive species. 
In this subsection, advances in ML for chemical engineering of 
plasma systems are explored. 
Zhu and co-authors\cite{zhuTailoringElectricField2022} have developed a DL framework called DeePlasKin for non-equilibrium plasma systems that can be described by a global chemistry model. The model is based on a predictor-corrector approach detailed in subsection\,\ref{ssec:control1}. For a given kinetic scheme and pre-defined temporal evolution of target species densities, it allows one to extract the temporal profile of the reduced electric field ($E/N$) and all other species densities. 
Pan and co-authors\cite{panDeepLearningassistedPulsed2023} have investigated a multi-layer feed-forward neural network for plasma kinetic modeling. 
The deep learning model has been applied to kinetic schemes for CH$_4-$Ar pulsed discharge and a N$_2-$H$_2$ plasma catalysis model for ammonia production. The magnitude of $E/N$ and time were chosen as the input data of the DNN, whereas the output was selected as the target densities of each species calculated at every instance of time by inquiring the kinetic model. The results showed that the network can replace complex kinetic models with large reactions set, while significantly improving the computational efficiency. 
Wang and co-authors\cite{wangModelingDischargeCharacteristics2023} have proposed an MLP
to describe the discharge characteristics and plasma chemistry of CO$_2$ pulsed DBD discharges at atmospheric pressure. 
The MLP was trained using data obtained from fluid simulations. As a result of their work, the trained model takes a few seconds to predict various features of 
CO$_2$ pulsed discharges, which significantly improves the computational efficiency with respect to the 30 hours CPU time that are required for the fluid model. 
Overall, these works have shown the great potential of ML for optimization and design of plasma sources in practical applications. 
In fact, the proposed models have been able not only to automatically extract plasma parameters that are difficult to measure in experiments, but also possess 
high computational efficiency, which provides a promising approach for multi-dimensional plasma modeling. 



\subsection{Plasma-surface interaction}
\label{ssec:psi1}

Particle fluxes onto LTP bounding surfaces contain a variety of species (e.g., electrons, neutrals, ions, radicals) which trigger different kinds of PSIs upon impact, for instance surface chemical reactions or sputtering. Each of which is more or less relevant for certain plasma applications, which are correspondingly engineered to lessen or strengthen one or the other mechanism. For modeling such processes and in particular the involved PSIs, a wide range of methodologies has been established, such as transport of ions in matter (TRIM)\cite{biersackMonteCarloComputer1980} or molecular dynamics (MD). Each method is limited by a combination and trade off between accuracy and computational feasibility. Recent advances in both regards utilizing ML will be outlined in the following for processes that either are driven by \emph{(i)} surface chemical reactions or \emph{(ii)} physical sputtering. 

\paragraph{i) Surface chemical reactions}

In addition to the species and surface chemical reactions inherent to ordinary gas-phase processing (e.g., chemical vapor deposition, atomic layer deposition, catalysis), plasmas introduce radicals, excited species, and ions and hereby enable a manifold of new reactions at the surface.
This circumstance is employed for instance for the plasma enhanced atomic layer deposition (PEALD) of metal oxides. It consists of two self-limiting half cycles with purging cycles in between, out of which the second half cycle, the plasma cycle, is dedicated to the deposition of O atoms, molecules, and radicals. Ding et al.\cite{dingMicroscopicModelingOptimal2020} have modeled PEALD of HfO$_2$ thin films with a kinetic Monte Carlo scheme. Reaction path ways and energy barriers were obtained by combining an intensive literature research with DFT calculations. Such simulations were used to assemble a data set to train a Bayesian regularized artificial neural network (BRANN)\cite{burdenBayesianRegularizationNeural2009}. The trained network was used for the prediction of the half cycle process times, which are essential for industrial considerations, as a function of total precursor pressure and surface temperature. The BRANN was constructed by combing an MLP network architecture with a Bayesian regularization, which penalizes large weights to achieve a better generalization.
In a follow-up work, \cite{dingMachineLearningbasedModeling2021} the authors pair their kinetic Monte Carlo model with a computational fluid dynamics process simulation. An RNN was then correspondingly trained to predict the evolution of the surface profile as well as gas-phase kinetics. Recurrent links establish internal states which allow for the prediction of the temporal dynamics. The issue of vanishing as well as exploding gradients during the training of the RNN was addressed by including an LSTM (cf. subsection\,\ref{ssec:concepts2}).

Plasma enhanced atomic layer etching (PEALE) of copper has been suggested by Sheil et al.\cite{sheilPreciseControlNanoscale2021} to include a half cycle of plasma induced oxidation followed by a half cycle of exposing the surface to etchant molecules (e.g., HCOOCH), which solely removes the just formed oxide layer. While the latter step is naturally self-limiting, the former is not. 
This finding has been seconded by Xia and Sautet,\cite{xiaPlasmaOxidationCopper2022} who performed MD simulations for a corresponding plasma oxidation of copper. A high-dimensional neural network potential was trained by considering the ReaxFF potential\cite{aktulgaParallelReactiveMolecular2012, zhuDevelopmentReactiveForce2020} and DFT calculations for the initial and on-the-fly data generation, respectively. The ML potential was demonstrated to overcome the limitations of classical as well as reactive interaction potentials imposed by their functional form, which are argued to not allow for accurate descriptions of very short range interatomic distances. 

For the particle fluxes from the plasma, not only radicals but molecules with high vibrational excitation are assumed to play an integral role for plasma catalysis. This is assessed although their contribution cannot be experimentally clarified due to an overlay with effects, for instance as induced by high electric fields. In contrast, simulations are well suited to provide experimentally unattainable insight.
Wan et al.\cite{wanDeepLearningAssistedInvestigation2022a} have conducted DFT simulations to investigate the role of electric field-dipole interactions for the CH$_3$ synthesis catalyzed with Ru surfaces. The resource intensive computations were accelerated by approximately five orders of magnitudes by generalizing the aggregated data with ANNs. Relevant atom configurations were processed with graph neural networks (GNNs),\cite{scarselliGraphNeuralNetwork2009, wuComprehensiveSurveyGraph2021} specifically graph convolutional networks,\cite{kipfSemiSupervisedClassificationGraph2022} whose network structure consists of nodes (i.e., atoms) and edges (i.e., configurational similarities). Multiple shared MLPs were used to combine the output with information on the electric field, adsorbate energy, and surface energy to predict all quantities of interest (i.e., adsorption energies, dipole moments, and polarizabilities of reactive intermediates).
Kedalo et al.\cite{kedaloApplicabilityFridmanMacheret2023} have performed MD simulations to show that the activation of highly excited N$_2$ molecules on Ru surfaces during NH$_3$ synthesis can be approximated by the Fridman--Macheret $\alpha$ model.\cite{fridmanPlasmaChemistry2008} The interatomic interactions were described by a ML potential, utilizing smooth overlap of the atomic positions (SOAP) descriptors,\cite{bartokRepresentingChemicalEnvironments2013} to enable ab initio accuracy but with increased computational efficiency.
Bal and Neyts\cite{balQuantifyingImpactVibrational2021} have demonstrated that the Fridman--Macheret $\alpha$ model is insufficient to describe the vibrational nonequilibrium (i.e., lowering of the free energy barrier) during the catalytic dissociation of H$_2$ and CH$_4$ on Ni surfaces at low temperatures of the order of 750 K. The multitudes of potential reaction pathways in between two metastable states were simplified by applying a harmonic linear discriminant analysis as a supervised dimensionality reduction algorithm.\cite{mendelsCollectiveVariablesLocal2018} The states themselves were studied by conducting short MD simulations with enhanced sampling methods (i.e., variationally enhanced sampling\cite{valssonVariationalApproachEnhanced2014}, metadynamics\cite{barducciWellTemperedMetadynamicsSmoothly2008, laioEscapingFreeenergyMinima2002}) to introduce the vibrational nonequilibrium. 

Siron et al.\cite{sironEnablingAutomatedHighthroughput2023} have proposed a high-throughput workflow for the study of amorphous material (i.e., a-C, a-Si, a-SiO$_2$, a-Al$_2$O$_3$) surface reactions relevant to dry plasma etching. Surface sites were characterized with SOAP descriptors and clustered with a Bayesian Gaussian Mixture model to effectively reduce sites to be studied for reaction kinetics, specifically etching. 

\paragraph{ii) Physical sputtering}

The bombardment of surfaces with energetic ions spawns collision cascades in the subsurface region, which eventually lead to persistent radiation damage and the emission of surface material atoms from the surface. The latter is referred to as sputtering and is of major importance for the sputter deposition of thin films as well as etching of high aspect ratio features.

Hamedani et al.\cite{hamedaniPrimaryRadiationDamage2021} have used an ML interaction potential to conduct an MD study on the radiation damage and sputtering of Si. The utilized ML potential, specifically the Gaussian approximation potential (GAP),\cite{bartokGaussianApproximationPotentials2010} was originally trained to equilibrium properties and, hence, complemented with a DFT repulsive potential (i.e., DMol\cite{nordlundRepulsiveInteratomicPotentials1997}) for short range interactions to resolve the evolution of the collision cascade accurately. The simulations provided sputtering yields that were in good agreement with experimental reference values. A greater number of isolated small defect clusters was observed than provided by MD simulations based on traditional interaction potentials (i.e., Stillinger-Weber\cite{stillingerComputerSimulationLocal1985}, Tersoff III\cite{tersoffNewEmpiricalApproach1988}). Moreover, a new defect structure was revealed, that is three split interstitials surrounding a central vacancy. 
This structure and the likelihood for its formation was further analyzed in a follow-up study.\cite{hamedaniInsightsPrimaryRadiation2020}  

Sputtering yields can also be predicted by utilizing the Yamamura model.\cite{yamamuraEnergyDependenceIonInduced1996} Phadke et al.\cite{phadkeSputterYieldsMonoatomic2022} have showed that its model parameter Q, which relates to linear scaling, has a simple positive correlation with the sputter threshold energy. This result was obtained by applying a Bayesian Markov Chain Monte Carlo algorithm to TRIDYN\cite{mollerTridynTRIMSimulation1984} simulations of Ar$^+$ and Ne$^+$ ions bombarding 70 elemental targets.
Eckstein and Preuss\cite{ecksteinNewFitFormulae2003} applied Bayesian probability theory to improve the accuracy of the Yamamura model's sputtering yield predictions by empirical means.   
Preuss et al.\cite{preussBayesianDeterminationParameters2019} conducted SDTrimSP\cite{mollerTridynTRIMSimulation1984, ecksteinSDTrimSPMonteCarloCode2007} simulations of D$^+$ ions sputtering Fe targets to revise the model's internal parameter (i.e., surface binding energy). This was achieved by to applying a reduced-order spectral expansion\cite{wienerHomogeneousChaos1938} uncertainty quantification based on the Bayesian framework.

\begin{figure}[tb!]
    \centering
    \includegraphics[width=8cm]{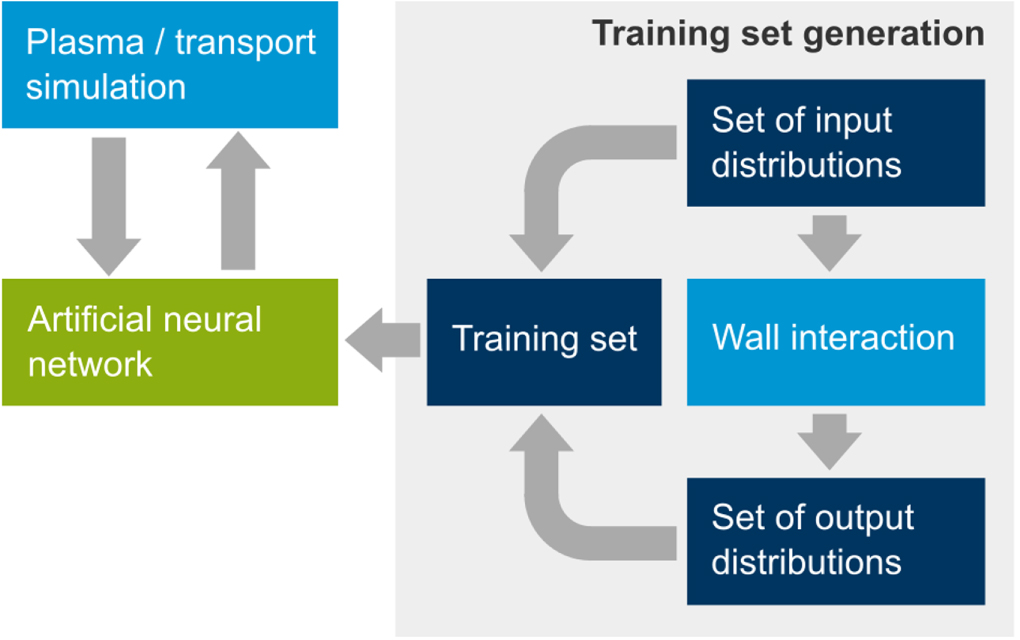}
    \caption{Conceptual diagram of the ANN plasma-surface interface model and the data flow for training set generation and run-time model evaluation. Figure reproduced from Krüger, Gergs, and Trieschmann\cite{krugerMachineLearningPlasmasurface2019} licensed under \href{http://creativecommons.org/licenses/by/4.0/}{CC BY 4.0}.}
    \label{fig:PSI_MLP}
\end{figure}

Krüger et al.\cite{krugerMachineLearningPlasmasurface2019} have conceptualized a PSI ML surrogate model for the Ar$^+$ ion bombardment of a TiAl composite target as indicated in Figure~\ref{fig:PSI_MLP}. The ML model, specifically an MLP, was meant to map the incident ions' energy distributions to the energy angular distributions (EADs) of sputtered particles. It was trained with noisy data from TRIDYN simulations (poor statistics), but demonstrated to be capable of predicting smooth EADs, which were almost congruent to reference TRIDYN simulations with high statistical quality.  
Gergs et al.\cite{gergsEfficientPlasmasurfaceInteraction2022} have continued those efforts by considering TiAl composite targets with varying compositions, introducing the stoichiometry as a basic system state to the model. A similar data aggregation scheme based on TRIDYN simulations was executed. The ML model was replaced by a combination of a convolutional $\beta$-variational autoencoder ($\beta$-VAE)\cite{kingmaAutoEncodingVariationalBayes2013, rezendeStochasticBackpropagationApproximate2014, higginsVVAELearningBasic2017} and CNN, defining a reduced dimensional (latent) space for the sputtered particles' EADs and mapping of the incident ion energy distributions as well as TiAl target composition to this particular latent space, respectively. Predictions and reference TRIDYN simulations agreed with each other, even though the assembled PSI ML surrogate model had got merely 0.39 \% of the MLP's\cite{krugerMachineLearningPlasmasurface2019} degrees of freedom and, hence, a significantly reduced likelihood for overfitting.
Kim et al.\cite{kimDeepNeuralNetworkbased2023} have pursued a similar but more advanced approach for the sputtering of fluorinated Si surfaces due to Ar$^+$ ion bombardment. MD instead of TRIDYN simulations were conducted to obtain information on the emitted (i.e., sputtered, reflected) particles' EADs as a function of the surface F coverage as well as Ar$^+$ ions' incident angle (but not kinetic energy). The aggregated EADs were used for a deep learning-based ROM (DL-ROM), comparing the performance of different ML architectures (i.e., autoencoder\cite{rumelhartLearningInternalRepresentations1985} (AE), denoising AE\cite{vincentExtractingComposingRobust2008}, VAE, conditional VAE (CVAE)\cite{sohnLearningStructuredOutput2015}) with each other. The CVAE was found to best its competitors. The PSI ML surrogate model was completed by introducing an additional MLP, which mapped the input information (i.e.,  surface F coverage, Ar$^+$ ions' incident angle) to the latent space established by the DL-ROM, enabling resource efficient predictions with atomic, MD fidelity.
Gergs et al.\cite{gergsPhysicsseparatingArtificialNeural2023} have performed hybrid MD/time-stamped force-bias Monte Carlo (tfMC)\cite{balTimeScaleAssociated2014, meesUniformacceptanceForcebiasMonte2012, neytsCombiningMolecularDynamics2013} simulations to setup a PSI ML surrogate model but treated sputtering of Al targets due to Ar$^+$ ion bombardment and growth of Al thin films at the substrate as generalized wall interactions. The incident particle flux composition as well as Ar$^+$ ion energy was varied to setup the data set.\cite{gergsMolecularDynamicsStudy2022} Two CVAEs were stacked to setup a physics-separating artificial neural network (PSNN), separating the ion bombardment induced damage from the damage to surface state (e.g., composition, mass density) translation. This network architecture has been argued to allow for training with experimental and simulation data in spite of an inherent discrepancy regarding the information accessibility (damage cannot be measured by similar means). The predictions mitigated the noise but maintained key physical features (e.g., threshold for Ar clustering, reflection probability) for small particle doses. 
In a follow-up work, \cite{gergsPhysicsseparatingArtificialNeural2023a} the authors have shared a methodology that enabled predictions for particle emission as well as growth during the reactive sputter deposition of AlN thin films in Ar/N$_2$ plasmas on experimental timescales, that is minutes to hours. First, the MD interaction potential was revised with an evolution strategy implemented in the genetic algorithm-based reactive force field optimizer method (GARFfield)\cite{jaramillo-boteroGeneralMultiobjectiveForce2014} to guarantee accurate PSI simulations.\cite{gergsChargeoptimizedManybodyInteraction2023} Second, a high-throughput, randomized data generation scheme was pursued to efficiently populate the relevant parameter space with hybrid MD/tfMC simulations.\cite{gergsPhysicsseparatingArtificialNeural2023a} Third, two CVAE were trained to predict either the PSIs or the diffusion processes. They were combined to form a PSNN\cite{gergsPhysicsseparatingArtificialNeural2023} as PSI ML surrogate model, which can readily be used to complement either plasma simulations or experimental diagnostics. The latter scenario was considered to successfully validate the model. Process times of 45~minutes were predicted within merely 34~GPU hours. In contrast, hybrid MD/tfMC simulations were argued to take more than approximately 8 million CPU years to finish on a comparable case study.

\subsection{Plasma process control and optimization}
\label{ssec:control1}

Many applications in LTP processing require precise control of the process and discharge parameters governing the deposition and/or removal of material from the plasma-facing surfaces (e.g., plasma etching). Whereas a detailed understanding through a virtual process simulation ideally provides all relevant intrinsic information in real-time, this pathway is often limited by a mismatch between the real process and its virtual replica, along with the typical run-time requirements of a process model. Therefore, several data-driven alternatives have been explored: \emph{(i)} MPC based on ROMs. \emph{(ii)} Plasma state estimation from incomplete observable information. \emph{(iii)} Data-driven process recipe design and optimization. \emph{(iv)} Data-driven parameter space exploration and optimization. Corresponding examples for these categories will be addressed in the following.

\paragraph{i) Model predictive control}
In a series of contributions, Woelfel et al. have developed MPC schemes for radio-frequency reactive magnetron sputtering fundamentally based on the Berg model.\cite{bergModelingReactiveSputtering1987} Following the identification of a ROM taking into account the nonlinear process dynamics due to surface poisoning,\cite{woelfelModelReductionIdentification2017} they have devised a controller for process operation at the transition between poisoned and clean surface modes using the previous reduced model, combined with measurements from multipole resonance probe experiments.\cite{woelfelModelApproximationStabilization2019, woelfelPlasmaStateControl2019} An improved reduced model and novel control scheme (also based on the Berg model) has been introduced later, taking into account the plasma state and its dynamical behavior.\cite{woelfelControlorientedPlasmaModeling2021}

The control problem of delivering a specific thermals dose during application of a kHz atmospheric pressure plasma jet (APPJ) operated in helium has been detailed by Mesbah and co-workers in a sequence of publications discussed in the following. Subsequent to model identification of a ROM (from experimental measurements), a feedback control strategy for 2D spatial thermal dose delivery was proposed.\cite{gidonPredictiveControl2D2019, gidonDatadrivenLPVModel2021} The substrate temperature as obtained from infrared thermal imaging was controlled based on the He flow and lateral shuttle position over a surface. The hierarchical feedback control strategy enabled reliable dose delivery despite abrupt disturbances that were introduced. An extension of the approach was suggested as treatment protocol, where the control of the surface temperature included the treatment time in an optimal control problem. One aspect was the adaptation of the control scheme to converge to the optimal solution for the real APPJ system.\cite{rodriguesDataDrivenAdaptiveOptimal2023}

The optimization of a plasma etch process has been investigated by Xiao et al. using a MPC scheme based on a physical multi-scale model pairing a 2D fluid description of the discharge and a kinetic Monte Carlo simulation of the surface kinetics. Initially, based on the simulated fluxes at the substrate surface, an RNN surrogate model was set up. Another one was implemented for the surface description, taking into account the stochasticity of the MC model.\cite{xiaoMultiscaleModelingRecurrent2021} Subsequently, a similar physical modeling approach has been proposed for a hierarchical surrogate model strategy.\cite{xiaoRecurrentNeuralNetworkBasedModel2022} Using similar 2D plasma fluid simulation results as for the previous approach, a ROM was devised via proper orthogonal decomposition and Galerkin's method.\cite{berkoozProperOrthogonalDecomposition1993} This reduced model was argued to capture the intrinsic system dynamics on a low-dimensional representation. It thereafter was used for training another RNN surrogate model for rapid prediction of the plasma discharge dynamics during argon-chlorine etching of silicon. In both works by Xiao et al.\cite{xiaoMultiscaleModelingRecurrent2021,xiaoRecurrentNeuralNetworkBasedModel2022}, MPC schemes have further been developed to control the etching depth and the feature bottom roughness of the process.

\paragraph{ii) Plasma state estimation}
With the goal to estimate the internal state of a LTP system in real time despite limited access to observable quantities, a physically-constrained extended Kalman filter (PC-EKF) has been devised by Greve et al.\cite{greveRealtimeStateEstimation2021} Firstly, the general principle including physical constraints (e.g., prohibiting negative quantities) was assessed at the example of the Lorenz system. The PC-EKF has been shown to closely resemble its chaotic system dynamics. Thereafter, the dynamics of a Hall effect thruster described by a global ionization model were considered. Based on particle balance equations for electrons and ions, the electron temperature was dynamically estimated based on the measured discharge currents. Successful recovery of the electron temperature was demonstrated by the investigation of four different discharge modes subject to varying types of ionization oscillations. The proposed scheme has been argued to benefit from improved robustness with the ability to estimate low-frequency dynamics in the presence of possibly not well resolved high-frequency dynamics (in the model or experiments). In a follow-up study also by Greve et al.,\cite{greveEstimationPlasmaProperties2022} the PC-EKF approach was generalized to more complex Ar and Ar/O$_2$ chemical reaction sets and global models of ICP sources. Particle balance equations as well as a power balance equation for electrons were taken into account. In an extensive case study, the robustness of the scheme has been demonstrated. Among other cases, a real time estimate of one or multiple unknown state variables based on corresponding limited measurements (e.g., electron absorbed input power based on argon ion density) has been shown, including pulsed operation.

\begin{figure*}[tb!]
    \centering
    \includegraphics[width=\textwidth]{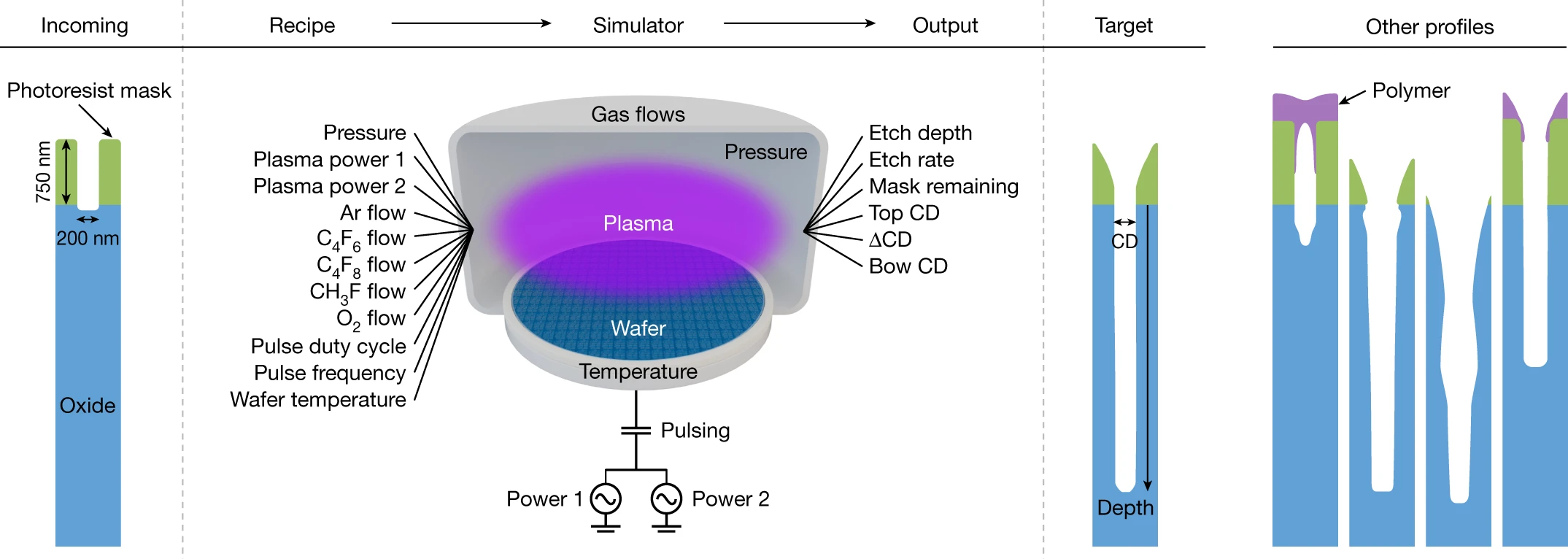}
    \caption{Schematic of the virtual process used for the process recipe design challenge. Figure reproduced from Kanarik, Osowiecki, Lu, Talukder, Roschewsky, Park, Kamon, Fried, and Gottscho\cite{kanarikHumanMachineCollaboration2023} licensed under \href{http://creativecommons.org/licenses/by/4.0/}{CC BY 4.0}.}
    \label{fig:virtual_process}
\end{figure*}

\paragraph{iii) Data-driven process recipe design and optimization}
In contrast to real time process control and estimation, a study by Kanarik et al.\cite{kanarikHumanMachineCollaboration2023} has investigated ML aided process recipe design. A virtual process of an radio-frequency etch plasma with fluorocarbon and O$_2$ chemistry coupled to a feature profile simulator to estimate etching and critical dimension (CD) has been used for reference. A schematic of the virtual process is depicted in Figure\,\ref{fig:virtual_process}. A comparison of junior and senior human experts against three computer algorithms to design a process recipe were evaluated, focusing on a minimization of the accumulated cost-to-target for achieving design goals. Their GPR model was found to greatly outperform the other algorithms (MCMC sampling, Tree-structured Parzen Estimator). Notably, a threshold was observed, determining the ideal handover from human to computer algorithm design for the specific challenge. This is linked to the amount of trials (and hence data samples) required for achieving the specified target. The proposed \emph{human first–computer last strategy} has been argued to reliably reduce the cost-to-target, whereas human experts and data-driven computer algorithms excel at different tasks (high vs low-dimensional parameter space exploration).

\paragraph{iv) Data-driven parameter space exploration and optimization}
Data-driven parameter space exploration in LTP has been predominantly pursued for atmospheric plasmas. As highlighted in subsection\,\ref{ssec:chemistry1}, Zhu et al.\cite{zhuTailoringElectricField2022} have investigated a global chemistry model (with the possibility to interface with a Boltzmann solver) for a DBD as a function of the electric field. A discrete time-stepping predictor/corrector scheme has been suggested: \emph{(a)} A reconstruction step to obtain a guess of the next reduced electric field $E/N$ value for a selected target density (at the next time step). \emph{(b)} A prediction step using an ordinary differential equation (ODE) solver of the physical model and the estimated $E/N$ to obtain the future system state. \emph{(c)} A correction step where the procedure is reevaluated if the difference between the outcome and the target exceeds a given threshold. The model enabled tailoring of the LTP system state evolution, allowing for efficient parameter space exploration and control of the plasma chemistry.

A more general approach to parameter space exploration has been suggested by Shao et al.\cite{shaoActiveLearningguidedExploration2022} based on GPR. The authors have trained an MLP as a surrogate model based on experimental data. Following the paradigm of Bayesian optimization (BO), this data-driven model is then exploited to set up a GPR model with an expected improvement acquisition function. For a considered use case of nitrogen fixation experiments of an atmospheric DC pin-to-pin glow discharge, for example, the energy cost of NO$_x$ generation was targeted for minimization. An active learning scheme based on this BO was established for exploration of a multi-dimensional parameter space.

\section{Advances}
\label{sec:advances}

The need for LTP and plasma chemistry databases have been previously highlighted (see section\,\ref{sec:review} and referenced literature).\cite{adamovich2022PlasmaRoadmap2022, anirudh2022ReviewDataDriven2023, alvesFoundationsPlasmaStandards2023} Many data-driven advances in modeling and simulation of LTP science and technology may be enabled only by a strategic data generation and acquisition initiative. This is particularly the case since data-driven techniques intrinsically require significant amounts of training data, which need to efficiently explore the relevant parameter space. While there is enormous potential in such data generation and exploitation, we reason that this is a matter of scientific practice and awareness but not a technical obstacle, to a substantial extent. In the following, in contrast, we focus on technical aspects that are hypothesized to have the potential to advance the field, proceeding from ML methods to LTP disciplines. We initially discuss ML concepts and methodology of general relevance to LTP modeling and simulation. Therein, we limit the discussion mainly to DL concepts. Thereafter, approaches and concepts specific to individual LTP problems in plasma physics, plasma chemistry, and plasma-surface interactions are addressed.

\subsection{ML methodology}
\label{ssec:concepts2}

Data-driven LTP modeling and simulation inherently addresses the description of complex physicochemical systems, on multiple dynamical time and length scales. Whereas some aspects of data-driven modeling targeting dynamical systems have been investigated in plasma science, a vast amount of concepts from related scientific disciplines appears to be under-explored in LTP. Therefore, in what follows we target widely applicable ML approaches to dynamical system modeling and their interpretation: \emph{(i)} Artificial neural network topologies. \emph{(ii)} Reduced order modeling and model identification. \emph{(iii)} Learning schemes for optimal data exploitation. \emph{(iv)} Explainable AI to support physics understanding. We aim to highlight the general concepts and draw the connection to specific aspects of plasma science in general and LTP processing in particular where applicable.

\paragraph{i) Artificial neural network topologies}
Several approaches to the modeling of dynamical systems using ANNs have been proposed. Early concepts have evolved around the class of RNNs.\cite{jaegerTutorialTrainingRecurrent2002} Recurrent network connections establish feedback loops between neurons, whereas the previous output of a neuron subject to a time delay is fed back to the same or another neuron. Feed-forward RNNs only maintain local recurrent connections per layer; fully recurrent network architectures establish a complete coupling between all neurons. Similar to ordinary fully-connected ANNs, recurrent connections may be multiplied with learnable weight parameters and passed through nonlinear activation functions. Training of an RNN involves the backpropagation of errors through time, passing them on through the evolution of the system.\cite{jaegerTutorialTrainingRecurrent2002} This might be challenging due to vanishing or exploding gradients and a large computational effort. Post training the (many) weight parameters of the RNN resemble a description of the system's equations of motion. Examples in LTP processing have already been addressed.\cite{xiaoMultiscaleModelingRecurrent2021, xiaoRecurrentNeuralNetworkBasedModel2022}

An important variant of RNNs are LSTM networks suggested by Schmidhuber and co-workers.\cite{hochreiterLongShortTermMemory1997, gersLearningForgetContinual2000} While similar in terms of recurrent feedback, their success is largely enabled by memory cells and gate units instead of basic neurons. Each memory cell with a self-connection includes gate units to control the information flow into and out of the cell, as well as of the recurrent feedback. As such, this may be attributed to the concepts of learning and forgetting,\cite{gersLearningForgetContinual2000} and allows for a substantially improved learning efficiency and robustness. LSTM networks have been applied in fluid dynamics for the prediction of the temporal dynamics of turbulent flow through channels based on direct numerical simulation (DNS) data.\cite{borrelliPredictingTemporalDynamics2022} They have further been used for the prediction of disruption in magnetic confinement fusion plasmas.\cite{zhengDisruptionPredictorBased2020, guoDisruptionPredictionEAST2021}

Moreover, in a study on a hierarchical approach to multiscale data-driven modeling, LSTM networks have been included for comparison.\cite{liuHierarchicalDeepLearning2022} Another fundamentally related variant of RNNs that was included in this study are echo-state networks (ESN).\cite{jaegerTutorialTrainingRecurrent2002, jaegerHarnessingNonlinearityPredicting2004} Therein, an RNN network \emph{reservoir} is initially set up. A number of inputs are fed into the RNN reservoir, while a number of outputs are extracted from the reservoir. In contrast to ordinary RNN networks, which comprise of learnable weight parameters at every neural node, only a subset of output weight parameters is adjusted during training. Consequently, a manifold of complex nonlinear system dynamics may inherently be evolved by the network, whereas only a small number of output weight parameters capture the desired system dynamics to be described by the data-driven model. To this end, a physics-informed ESN architecture has been investigated for the description of chaotic systems.\cite{doanPhysicsinformedEchoState2020} For the chaotic dynamics of a Lorenz system and a Charney--DeVore system, robust training was achieved with an extended prediction time horizon of approximately two Lyapunov times.

Following the physics-informed learning paradigm,\cite{raissiPhysicsinformedNeuralNetworks2019} the amount and the variants of approaches exceeds the frame of what can be covered here. The interested reader is referred to an extensive review on PINNs in scientific ML by Cuomo et al. (which also partially covers RNNs and its variants).\cite{cuomoScientificMachineLearning2022} 
It may be noted that the technical implementation of PINNs is subject to ongoing research. This is partially due to the limited performance and convergence characteristics of PINNs in general, and of ordinary ANNs such as an MLP in particular.\cite{wangUnderstandingMitigatingGradient2021,wangNASPINNNeuralArchitecture2024}
Another approach to physics-informed ML extending this idea is based on deep neural operator networks (DeepONets).\cite{luLearningNonlinearOperators2021} With the reasoning that ANNs can be considered universal operator approximators (similar to the property that ANNs are universal function approximators), these networks comprise of a trunk network and a (single or stack of) branch network(s). The branch part takes as input discrete \emph{sensor} samples $u(x_i)$ of a function $u(x)$. The trunk part takes as input the quantity $y$. The output of both sub-networks is combined to provide a functional representation $G(u)(y)$. Following an offline training for a variety of examples, Lu et al.\cite{luLearningNonlinearOperators2021} have demonstrated online inference performance significantly improving on other reference ANNs (among others LSTM networks). Different variants and extension taking advantage of Fourier and Laplace transformations as nonlinear kernels have further been proposed.\cite{liNeuralOperatorGraph2020,liFourierNeuralOperator2021, caoLNOLaplaceNeural2023}

Transformer architectures have been originally developed in NLP.\cite{vaswaniAttentionAllYou2017} The examples previously addressed have fundamentally explored similar architectures without substantial conceptual modifications (e.g., either by using standalone transformer layers\cite{kawaguchiPhysicsinformedNeuralNetworks2022,kawaguchiDatadrivenDiscoveryElectron2023,kimNumericalStrategySolving2023} or similar to an image-to-image translation task\cite{siffaMachineLearnedPoissonSolver2023}). Classical transformer networks process input information through embeddings that transform the input data into a corresponding vector space representation. Recently, an extension to the concept of transformer networks has been suggested by Geneva and Zabaras for surrogate modeling of physical systems.\cite{genevaTransformersModelingPhysical2022} Therein, alternative embeddings have been suggested based on Koopman dynamics in a two step learning procedure. Initially, embedding ANNs are trained in an encoder-decoder structure to approximate the Koopman based dynamics in a latent space representation. Thereafter, a transformer network is trained using the pretrained Koopman embeddings. The obtained data-driven surrogate models have been demonstrated to effectively capture the chaotic system dynamics of a Lorenz system, the 2D fluid dynamics of a transient flow, and 3D reaction-diffusion dynamics. In comparison the Koopman based transformer models have been demonstrated to outperform alternative surrogate models of such dynamical physical systems (including other embeddings and LSTM networks).

\paragraph{ii) Reduced order modeling and model identification}
Reduced-order models have previously been used in LTP modeling and simulation as exemplified above. The particular notion of sparse identification of nonlinear dynamics (SINDy),\cite{bruntonDiscoveringGoverningEquations2016, rudyDatadrivenDiscoveryPartial2017} has been demonstrated to provide interpretable and generalizable nonlinear differential equations of reduced order which effectively capture the dynamics of the full model. This is achieved by identifying the relevant system dynamics from a library of potential functional forms by means of a sparse regression. It has been applied to space and astrophysical plasmas based on PIC simulations by Alves and Fiuza demonstrating the potential of recovering the fundamental dynamics of the high-fidelity kinetic plasma simulations.\cite{alvesDatadrivenDiscoveryReduced2022} It may be straightforwardly applicable to LTP modeling and simulation data to capture their nonlinear dynamics, albeit a complex chemistry may necessitate further adaptation. Sparse regression has similarly been applied by Kaptanoglu and co-workers in the context of magnetic confinement fusion plasmas for the optimization of magnetic fields in stellarator fusion plasmas.\cite{kaptanogluPermanentMagnetOptimizationStellarators2022, kaptanogluSparseRegressionPlasma2023} Moreover, MPC based on such ROMs has been suggested with limited availability of training data.\cite{kaiserSparseIdentificationNonlinear2018} An aspect that has been applied using time-dependent simulations of the tokamak plasma boundary.\cite{loreTimedependentSOLPSITERSimulations2023, kaptanogluSparseRegressionPlasma2023} It may be specifically relevant also in the context of LTP processing with its diverse requirements for precise process control.

\paragraph{iii) Learning schemes}
In contrast to classical supervised learning schemes, alternative approaches may be relevant also due to limited availability of training data. In a Bayesian setting, provided with an estimate of the prediction uncertainty, active learning has been exploited for efficient parameter space exploration.\cite{settlesActiveLearningLiterature2009} While such an uncertainty metric is straightforward in the context of GPR modeling,\cite{shaoActiveLearningguidedExploration2022} similar techniques have been adapted to MD simulations and active learning of corresponding ML surrogate models.\cite{diawMultiscaleSimulationPlasma2020} 

Transfer learning enables an alternative route to ML in the small-data limit.\cite{bozinovskiReminderFirstPaper2020} The fundamental idea relates to the property of generalization that data-driven models pursue. Given an initial problem task with abundant data, an ML model may be robustly trained. The pretrained model may be subsequently retrained on a related, yet different small-data task, whereas just a minor subset of weight parameters are allowed to be adjusted. Hence, the robustness and generalized prediction of the initial model may be successfully transferred to the more challenging learning task. This approach has been successfully applied for instance in the design optimization in inertial confinement fusion.\cite{humbirdTransferLearningDriven2022}

\paragraph{iv) Explainable AI}
As surveyed by Gilpin et al.\cite{gilpinExplainingExplanationsOverview2018}, a multitude of explainable AI (XAI) concepts have been proposed for the interpretation and explanation of data-driven models. As an example, local interpretable model-agnostic explanations (LIME) has been suggested for identifying interpretable models that locally approximate the nonlinear behavior of any underlying model.\cite{ribeiroWhyShouldTrust2016} While this is possibly an underexplored aspect in the context of LTP, an example where LIME relates to the data-driven classification and interpretation of OES experiments of plasma in aqueous solution has been proposed.\cite{wangMachineLearningExplainable2021} Further works by Lundberg et al.\cite{lundbergUnifiedApproachInterpreting2017} has included LIME and other methods in an extension Shapley additive explanations (SHAP). It is similarly based on local explanation models, but defines additive feature attribution methods to unify several of the included interpretable methods. There is arguably no mathematical proof that physics-based and physics-informed ML, and corresponding models capture exactly the desired dynamics (and only those). LTP modeling and simulation has got its inherent foundation in a rigorous mathematical description and numerical solution. XAI methods may not only foster advances in data-driven approaches to the field, but strongly contribute to the general acceptance in LTP science and technology.

\subsection{Plasma Physics}
\label{ssec:discharge2}
Modeling and simulation of LTP physics is traditionally facing two substantial challenges. Firstly, their complex system dynamics typically involve multiple physical phenomena and, secondly, these dynamics spread across many orders of magnitude in length and time. As previously outlined, both challenges may be addressed and partially solved by i) data-driven surrogate modeling of sub-problems and ii) data-driven surrogate modeling of the complete discharge dynamics. Despite intrinsic progress in the LTP field alone, these efforts may be largely supported through consideration of advances in adjacent research disciplines, as discussed in the following.

\paragraph{i) Data-driven surrogate sub-modeling}
\begin{figure*}[tb!]
    \centering
    \includegraphics[width=\textwidth]{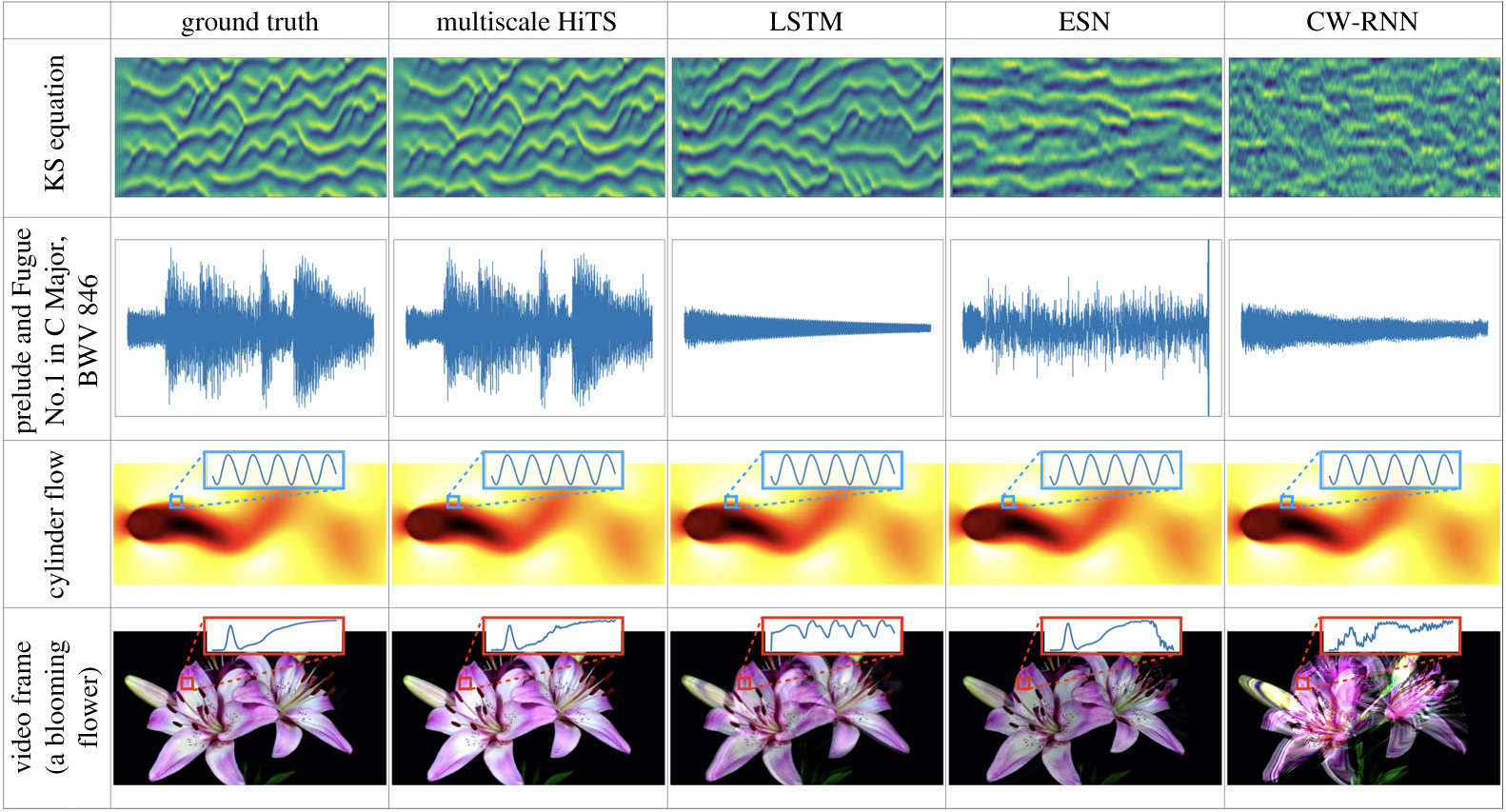}
    \caption{Output of different ANN architectures for the prediction of different training sequences (Kuramoto--Sivashinsky equation, music, flow past a cylinder, video). Figure reproduced from Liu , Kutz, and Brunton\cite{liuHierarchicalDeepLearning2022} licensed under \href{http://creativecommons.org/licenses/by/4.0/}{CC BY 4.0}.}
    \label{fig:hierarchical_stepping}
\end{figure*}
Describing the evolution of dynamical systems (or sub-systems) accurately is an omnipresent task in scientific computing. The time integration schemes of differential operators have been correspondingly considered in different fields (e.g., pairing Runge--Kutta integration schemes with ANNs \cite{wangRungeKuttaNeuralNetwork1998, zhongLowtemperaturePlasmaSimulation2022, liuHierarchicalDeepLearning2022, zhuConvolutionalNeuralNetworks2023}). For the time integration of multiscale physical systems, Liu et al.\cite{liuHierarchicalDeepLearning2022} have suggested an hierarchical time-steppers (HiTSs) framework. The approach builds on the decomposition of the system dynamics into a hierarchy of prediction steps that can be arranged and interpolated in time. It has been shown to outperform other methods based on LSTM, ESN, and clockwork RNNs for the Lorenz system, 2D fluid flow past a cylinder, the Kuramoto--Sivashinsky (KS) equation, as well as music and video sequences. An illustrative example comparison of the prediction output is given in Figure\,\ref{fig:hierarchical_stepping}. LTP modeling and simulation is facing essentially alike problems, suggesting it may similarly benefit from application of the HiTSs method (and/or other methods referenced) in the description of their multiscale dynamics.

An approach using the concept of dynamic mode decomposition (DMD) to obtain low-dimensional linear dynamics of given data has been applied to magnetic confinement fusion plasma simulations by De Pascuale et al.\cite{depascualeDatadrivenLinearTime2022} The obtained dynamic modes are evolved individually solving a linear ODE. The constructed linear time advance operator based on DMD has been successfully applied for the analysis of several 1D diffusion transport problems, as well as for the dynamics of 1D profiles at the divertor targets of a tokamak plasma. This scheme could be similarly applied to complex LTP processes.

The solution to Newton's equations for particle based simulations has been explored by Kadupitiya et al. using LSTM networks.\cite{kadupitiyaSolvingNewtonEquations2022} The accurate applicability has been investigated for few-particle systems subject to various interaction potentials (e.g., Lennard--Jones). Energy-conserving dynamics up to 4000$\times$ larger timesteps compared to a Verlet integration have been demonstrated. Following the concept of GNNs,\cite{scarselliGraphNeuralNetwork2009, wuComprehensiveSurveyGraph2021} which embed the interaction dynamics of nodes (e.g., particles) in their learnable edge weights (e.g., interaction), the integration of ODEs in an Hamiltonian framework, have been discussed by Sanchez-Gonzalez et al.\cite{sanchez-gonzalezGraphNetworksLearnable2018, sanchez-gonzalezHamiltonianGraphNetworks2019} The proposed Hamiltonian ODE graph network trained on Runge--Kutta scheme data generalizes the integrators to accurately represent the Hamiltonian dynamics. In the context of many-particle simulations, Li et al.\cite{liGraphNeuralNetworks2022} have proposed a GNN accelerated MD framework that predicts the interaction forces for the system, given its momentary state (e.g., atom positions and types). It has been argued that the scheme can be easily integrated into existing numerical integrators. Due to the consideration of multi-particle interactions, this scheme may be specifically relevant in PIC simulations should Coulomb interactions be important.

\paragraph{ii) Data-driven discharge surrogate modeling}

Many of the ML methods that may be transferred to data driven surrogate modeling of LTPs were discussed in subsection\,\ref{ssec:concepts2}. This relates specifically to the representation of dynamical systems through recurrent ANN models. A few examples from plasma science and adjacent fields were covered. However, it goes beyond the scope of this paper to cover related works from fluid dynamics. The reader is referred to corresponding comprehensive surveys by Brunton and Vinuesa et al.\cite{bruntonMachineLearningFluid2020, vinuesaEnhancingComputationalFluid2022a}

Concerning data-driven process control and MPC it should be noted that the plasma information based VM methodology for LTP etching processes previously introduced may be adapted based on physical or data-driven model data.\cite{maggipintoDeepVMDeepLearningbased2019, parkPredictiveControlPlasma2020, parkMicrorangeUniformityControl2021} Despite a possible discrepancy between models and reality, in particular systematic deviations may be straightforwardly taken into account in a data-driven approach by means of discrepancy learning. A learnable translation layer may suffice to capture and combine the relevant information by means of data-fusion. This may enable data-driven MPC with limited observables (cf.\ also PC-EKF\cite{greveRealtimeStateEstimation2021,greveEstimationPlasmaProperties2022}). Additionally, MPC schemes based on LSTM networks that have been successfully demonstrated in magnetic confinement fusion plasmas subject to inherent complex nonlinear dynamics by Degrave et al.\cite{degraveMagneticControlTokamak2022} may be promising candidates for adaptation in LTP processing.

\subsection{Plasma chemistry}
\label{ssec:chemistry2}

Several challenges arise in modeling and simulations of plasma chemistry. 
These are linked to the setup of accurate reaction mechanisms and high computational cost of plasma models with complex chemistries. 
This subsection describes advances in data-driven models and applications for such systems. 
Examples are mainly taken from the field of combustion, reactive gas flow modeling, and astrophysics. 
In particular, two different topics are discussed for i) dimensionality reduction of complex chemical networks and 
ii) data-driven approaches for fast and accurate chemical models. 
Such topics offer promising solutions for future investigations that are focused on bridging the gap between low cost and high accuracy in plasma chemistry models.

\paragraph{i) Reduction of complex chemical networks}

The adoption of detailed mechanisms for chemical kinetics often poses two types of challenges. 
First, the number of degrees of freedom is large; and second, the dynamics is characterized by a wide range of time scales.
This has motivated the development of several techniques for reducing the complexity of such kinetic models where only a few variables are considered in the simplified model.
Grassi and co-authors\cite{grassiReducingComplexityChemical2022} have used AEs to reduce complex chemical data sets and decrease the computational time associated with numerical solutions of ODEs for a large number of species and reactions.
In particular, an encoder is used to reduce the dimensionality of the chemical state space into a latent (lower-dimensional) space.
Analogously to the chemical network in the original space, that is represented
by the system of ODEs, the chemical network in the latent space can be evolved in time with a different set of ODEs which has got a significantly smaller number of dimensions.
Hence, a decoder is used to transform the variable from the latent space back to the original space.
The proposed model enables a compression of chemical networks composed of 224 reactions and 29 species into a network with 12 reactions and 5 species that are evolved in time using a standard ODE solver. Their work has demonstrated that a 65-times speed-up can be achieved. 
Tang and Turk \cite{tangReducedOrderModel2022} have proposed a combination of AE and neural ODE to model the temporal evolution of chemical kinetics in a reduced subspace. The architecture of the network is similar to the one by Grassi and co-authors \cite{grassiReducingComplexityChemical2022}. 
In their work, a 10-fold speed-up compared to commonly used astro-chemistry solver for a 9-species primordial network  has been achieved, while maintaining $1\%$ accuracy across a wide range of density and temperature.
As described in subsection \ref{ssec:chemistry1}, it is well known that, due to the presence of fast and slow dynamics, the chemical systems are characterized by low-dimensional manifolds in the concentration space. Chiavazzo and co-authors\cite{chiavazzoReducedModelsChemical2014} applied the Diffusion Map (DMAP) 
approach for constructing reduced kinetics models for combustion applications. The DMAP approach can be considered as a non-linear counterpart of PCA that can be used for searching a low-dimensional embedding of a high-dimensional chemical data set.

\paragraph{ii) Fast and accurate models of complex chemistries}

Stiff equations for chemical kinetics are computationally expensive to solve numerically. 
The efficacy of ML for applications that need to solve systems of ODEs or PDEs has been explored in many areas, such as computational fluid dynamics (CFD) and combustion.
Goswami and co-authors\cite{goswamiLearningStiffChemical2023} have developed a neural operator-based surrogate model (DeepONet) to efficiently
solve stiff chemical kinetics. The model was applied to a chemical reaction system for CO/H$_2$ burning of syngas, which contains 11 species and 21 reactions,
and a temporally developing planar CO/H$_2$ jet flame (turbulent flame) using the same syngas mechanism. 
It has been shown that the DeepONet, once trained, can accurately integrate the thermochemical state for
arbitrarily large time advancements, leading to significant computational gains compared to stiff integration schemes.
Moreover, both results obtained with DeepONet and AE-based DeepONet are computationally very efficient compared to conventional CFD solvers.
One of the advantages of this approach is the possibility to be included in a CFD solver for solving such stiff
chemically reacting problems, which can drastically reduce the computational cost.
Campoli and co-authors\cite{campoliAssessmentMachineLearning2022} have performed state-to-state numerical simulations of high speed 
reacting gas flow using data-driven ML regression models. 
The models were applied to a system of equations for a 1D reacting flow of a five-component air mixture including a total of 122 excited states of N$_2$, O$_2$, NO, N, and O species with detailed vibrational kinetics. Several methods have been compared for estimation of the state-to-state relaxation rate terms, such as kernel ridge (KR), SVM, k-nearest neighbor, Gaussian
processes, decision tree (DT), random forest, extremely randomized trees, gradient boosting, histogram-based gradient boosting, and MLP. From their comparison, it has been found that KR reports the lowest error levels while the SVM reports the highest. Comparable error levels are reported by the remaining algorithms; however, there are noticeable differences in the prediction time, where DT is found to be the
fastest algorithm.
Finally, the solution of the state-to-state Euler system of equations was inferred by means of a deep neural
network by-passing the use of the solver. As a result of this study, it has been been demonstrated that embedding of ML algorithms into ODE solvers offers a speed-up of several orders of magnitude.
Chemical reactor networks (CRNs) have been investigated by Savarese and co-authors \cite{savareseMachineLearningClustering2023} for performing faster 
CFD simulations of combustion systems. In their work, a novel automatic data-driven method for the design of CRN models has been proposed. 
The method is based on a combination of unsupervised clustering and graph scanning algorithms.
Results of their work show that the novel methodology is capable of extracting equivalent CRN models of realistic combustion devices from CFD data. 
Furthermore, the unsupervised clustering algorithm was able to automatically detect important regions within the combustor domain, such as inlets, inflame and post-flame reactors. 
Overall, the approach is promising as it has the potential to drive towards the development of ROMs without a huge amount of user-based knowledge. 
Mao and co-authors\cite{maoDeepFlameDeepLearning2022} have developed a novel CFD solver that combines the multi-functionalities of the OpenFOAM library\cite{OpenFOAMFoundationLtd2023} with the ML framework Torch\cite{collobertTorch7MatlablikeEnvironment2011} and chemical kinetics program Cantera\cite{goodwinCanteraUserGuide2002}. 
In their solver, DNNs are used to accelerate chemistry solvers and improve the simulation efficiency. As a result, a speed-up of two orders of magnitude is achieved
in a simple hydrogen ignition case when performed on a medium-end GPU. This work highlights the potential of the integration of ML techniques within current libraries and computing architecture for detailed CFD simulations of complex hydrocarbon fuels. 

\subsection{Plasma-surface interaction}

The review of ML utilized for PSIs was subdivided by means of physics (i.e., surface chemical reactions, sputtering) as well as related processes (e.g., PEALD/PEALE), highlighting differences and how pioneering works utilized ML to advance the research in such fields. The consideration of ML methods established and proposed for similar processes without plasmas (e.g., thermal atomic layer deposition/etching) may allow one to adopt novel techniques as well as inspire new studies. This opportunity is particularly valuable for research on plasma catalysis, where ML has already been applied in manifold ways and promising advances are foreshadowed. An excellent overview on ML assisted modeling and simulations of heterogeneous catalysis has been put together by Mou et al.\cite{mouBridgingComplexityGap2023}, whereas Chen et al.\cite{chenMachinelearningAtomicSimulation2023} have assembled a more detailed review for the most recent accomplishments due to ML potential-based atomic simulation. 

In the following, overarching challenges relevant to all fields of LTP processing are discussed.
The intrinsic time and length scales of the two states of matter (i.e., plasma, solid-state) differ in orders of magnitude. From a material scientist point of view, who resolves atomic time and length scales, PSIs have to be perceived as individual or a sequence of impinging particles, neglecting the true spatiotemporal correlations. Multiscale modeling approaches which repeatedly substitute short time and length scales are, however, always accompanied with an information loss of the fundamental atomic phenomena. ML is assumed to provide an alternative. 

For electrons, information on their dynamics, specifically, on the secondary electron emissions are essential for LTP modeling and simulation.\cite{phelpsColdcathodeDischargesBreakdown1999} Although experimental as well as theoretical approaches have been pursued in the past, the availability of such coefficients is scarce. More recent efforts include the combination of both (i.e., PIC/MCC simulations, phase resolved OES) and was framed as $\gamma$-CAST.\cite{dakshaComputationallyAssistedSpectroscopic2016, schulzeComputationallyAssistedTechnique2022} Moreover, a review of Monte Carlo methods to simulate the electron scattering in solids was put together by Chang et al.,\cite{changCalculationSecondaryElectron2018} who studied the low-energy electron deposition in W surfaces. Electron force field simulations have been demonstrated to predict electron
emissions due to the Auger process.\cite{suMechanismsAugerinducedChemistry2009}. Auger neutralization was also described by Pamperin et al.,\cite{pamperinIoninducedSecondaryElectron2018} who have formulated a generic quantum-kinetic approach, that is based on Anderson-Newns-type effective Hamiltonians. Bronold and Fehske\cite{bronoldInvariantEmbeddingApproach2022} have derived an equation for the electron emission yield in case of low energy electrons impinging onto metal surfaces. An invariant embedding principle was employed. It is argued that such approaches or combinations thereof could synergistically complement each other, in particular when paired with corresponding ML methods.

Information on the atomic scale have been limited for a long time by a compromise of accuracy and computational efficiency, effectively restricting the physical fidelity of related case studies. Recent advances by means of ML interaction potentials have indicated to eventually overcome this dilemma. However, only a small number of such works have been published so far in the context of LTP modeling and simulation (cf. subsection\,\ref{ssec:psi1}). It is therefore essential to further encourage their usage, enabling more likely developments that meet demands specific to plasma processing. The burden of setting up suitable ML interaction potentials could be bypassed at least to some extent by joining available ones with more traditional methods. For example, the Ziegler--Biersack--Littmark potential\cite{zieglerStoppingRangeIons1985a} and all-electron DFT repulsive potential, DMol, have been merged with ML interaction potentials (e.g., GAP), which were trained to equilibrium properties, to account for screened nuclear repulsion during high-energy collisions when radiating Si or Al.\cite{wangDeepLearningInteratomic2019, hamedaniInsightsPrimaryRadiation2020, niuMachinelearningInteratomicPotential2023, hamedaniPrimaryRadiationDamage2021} Caro et al.\cite{caroGrowthMechanismOrigin2018,caroMachineLearningDriven2020} conducted MD simulations with the GAP to study the ion beam deposition of C thin films. Studying a complementary sputter deposition can be readily achieved by introducing Ar$^+$ ions to the process, whose interactions are sufficiently well described by traditional means. An overview of and introduction to ML interaction potentials for the popular Large-scale Atomic/Molecular Massively Parallel Simulator (LAMMPS) has been presented by Thompson et al.\cite{thompsonLAMMPSFlexibleSimulation2022}
    
The increased efficiency of such approaches allows for a more thorough but still accurate parameter space exploration. This is of major importance since most PSIs (e.g., surface chemical reactions, ion radiation) are subject to excessive degrees of freedom and cannot straightforwardly be evaluated (e.g., reaction path ways, 0D to 3D defect structures). Remedies combine ML algorithms for clustering (e.g., Bayesian Gaussian mixture model), dimensionality reduction (e.g., harmonic linear discriminant analysis), or surrogate modeling (e.g., PSNN) with high-throughput simulations (e.g., randomized trajectories,\cite{gergsPhysicsseparatingArtificialNeural2023a} amorphous surface site melt-quenching\cite{sironEnablingAutomatedHighthroughput2023}) or enhanced sampling methods (e.g., variationally enhanced sampling, metadynamics).\cite{limEvolutionMetastableStructures2020, balQuantifyingImpactVibrational2021, sironEnablingAutomatedHighthroughput2023, gergsPhysicsseparatingArtificialNeural2023a}. Notably, GNN have also been demonstrated to significantly accelerate atomic simulations.\cite{wanDeepLearningAssistedInvestigation2022a} An active learning scheme as applied by Diaw et al.\cite{diawMultiscaleSimulationPlasma2020} when conducting MD simulations of warm dense matter to setup a corresponding surrogate model may be considered to secure spending available resources most effectively.

Whereas samples in the parameter space are naturally related to different process conditions, they may also be viewed as instances or states of a system (surface patch), which may evolve during operation.
Lim et al.\cite{limEvolutionMetastableStructures2020} formulated an according transition state modeling approach\cite{henkelmanDimerMethodFinding1999, henkelmanClimbingImageNudged2000, henkelmanImprovedTangentEstimate2000} for the long-time scale (i.e., of the order of microseconds) restructuring of Pd deposited on Ag. States were found by detecting relevant positional changes of adatoms during MD simulations, clustering them (unsupervised), and eventually sorting them (supervised). 
Gergs et al.\cite{gergsPhysicsseparatingArtificialNeural2023a} used a PSNN to generalize on previously sampled but disjoined system state trajectories. The system alternately underwent PSIs and diffusion processes for given experimental discharge conditions, continuously evolving the system state (e.g., composition, point defect populations). Following a similar line of thought, single instances or states could also be understood as surface patches, which may be assembled to setup a PSI ML surrogate model on the feature scale. 
Kim et al.\cite{kimDeepNeuralNetworkbased2023} suggest a similar usage for etching high-aspect ratio feature patterns. However, such an approach may also be of significant interest for research on plasma catalysis, taking complex large-scale surfaces into account.
Joining such ML models, which are created in a bottom-up style (starting with simulations on the microscopic scale), with a complementary top-down approach (starting with experiments on the macroscopic scale) may ultimately enable a fidelity beyond DFT.

\section{Concluding remarks}
\label{sec:conclusion}
Scientific ML has had a significant impact in many research disciplines. In modeling and simulation of LTPs, it has experienced increased interest and important developments primarily within the past few years. In this time period, however, already a substantial interdisciplinary exchange could be witnessed. Within this review we had two main objectives: \emph{(a)} Review the state of the art in ML and data-driven approaches to LTP modeling and simulation. With the goal to highlight specific use cases and applications, we divided our survey by physicochemical and modeling aspects relevant for LTP processing (rather than methodologies). \emph{(b)} Provide a perspective of potential advances. These may be enabled through transfer and inspiration of current methodologies from other scientific disciplines as well as future ML developments in general. It certainly demands adaptation to the requirements faced in LTP processing, e.g., including the system dynamics for a precise control of deposition/etching process at the atomic level.

Open access to literature, databases, research data and codes, as well as data-driven models are foreseen to be an integral part in these efforts.\cite{frankePlasmaMDSMetadataSchema2020, carboneDataNeedsModeling2021, siffaAdamantJSONSchemabased2022a, alvesFoundationsPlasmaStandards2023}
Sharing information following the findable, accessible, interoperable, and reusable (FAIR) principles\cite{wilkinsonFAIRGuidingPrinciples2016} would allow researchers independent of status and financial budget to develop the field. We believe this may enable vital new pathways for research on LTP processing.

It is important to stress that taking advantage of these developments should be assessed on a rational basis. A (more or less) critical evaluation should therefore ask which advances are truly enabled by corresponding ML methods. An answer may relate simply to the run-time performance of a corresponding numerical simulation. Moreover, it may pertain novel insights previously hidden within data, e.g., results from numerical simulation. ML methods may ultimately uncover unprecedented insights and interpretation otherwise not attainable. An associated aspect further relates to the net computational budget required to obtain certain data sets, the training procedure, the costs of run-time evaluation, and the capability of generalization of the data-driven models. This may be contrasted with the effort of evaluating individual numerical simulations.

Despite being generally true for science and technology, it may be envisioned that scientific ML and data-driven approaches to LTP modeling and simulation may not only support the analysis of known unknowns. In contrast, due to its inherent propensity to spotlight hidden patterns in data, it may encourage discovery of unknown unknowns.
This aspect is argued to offer unprecedented potential for advances in plasma science and technology and beyond.

\section*{Data and Code Availability Statement}
Data sharing is not applicable to this article, as no new data were created or analyzed.

\section*{Acknowledgment}
Funded by the Deutsche Forschungsgemeinschaft (DFG, German Research Foundation) -- Project-ID 138690629 (TRR 87), Project-ID 434434223 (SFB 1461), and Project-ID 445072286.

\section*{ORCID}
\noindent
J. Trieschmann: \url{https://orcid.org/0000-0001-9136-8019} \\
L. Vialetto: \url{https://orcid.org/0000-0003-3802-8001} \\
T. Gergs: \url{https://orcid.org/0000-0001-5041-2941}


\bibliography{references}   

\begin{thebibliography}{100}

\bibitem{gottschoMicroscopicUniformityPlasma1992}
R.~A. Gottscho, C.~W. Jurgensen, and D.~J. Vitkavage, ``Microscopic uniformity
  in plasma etching,'' {\em Journal of Vacuum Science \& Technology B:
  Microelectronics and Nanometer Structures Processing, Measurement, and
  Phenomena} {\bf 10}, 2133--2147  (1992).

\bibitem{oehrleinFoundationsLowtemperaturePlasma2018}
G.~S. Oehrlein and S.~Hamaguchi, ``Foundations of low-temperature plasma
  enhanced materials synthesis and etching,'' {\em Plasma Sources Science and
  Technology} {\bf 27}(2), 023001  (2018).

\bibitem{tomieTinLaserproducedPlasma2012}
T.~Tomie, ``Tin laser-produced plasma as the light source for extreme
  ultraviolet lithography high-volume manufacturing: History, ideal plasma,
  present status, and prospects,'' {\em Journal of Micro/Nanolithography, MEMS,
  and MOEMS} {\bf 11}, 021109  (2012).

\bibitem{bishopNeuralNetworksPattern1996}
C.~M. Bishop, {\em Neural {{Networks}} for {{Pattern Recognition}}}, {Oxford
  University Press}, {Oxford, UK}  (1996).

\bibitem{rasmussenGaussianProcessesMachine2005}
C.~E. Rasmussen and C.~K.~I. Williams, {\em Gaussian {{Processes}} for
  {{Machine Learning}}}, {MIT Press}  (2005).

\bibitem{haykinNeuralNetworksLearning2008}
S.~Haykin, {\em Neural {{Networks}} and {{Learning Machines}}: {{A
  Comprehensive Foundation}}}, {Prentice Hall International}, {New York, USA},
  3rd~ed.  (2008).

\bibitem{goodfellowDeepLearning2016}
I.~Goodfellow, Y.~Bengio, and A.~Courville, {\em Deep {{Learning}}}, {MIT
  Press}, {Cambridge, USA}  (2016).

\bibitem{geronHandsOnMachineLearning2022}
A.~G{\'e}ron, {\em Hands-{{On Machine Learning}} with {{Scikit-Learn}},
  {{Keras}}, and {{TensorFlow}}: {{Concepts}}, {{Tools}}, and {{Techniques}} to
  {{Build Intelligent Systems}}}, {O'Reilly Media}, 3rd~ed.  (2022).

\bibitem{adamovich2022PlasmaRoadmap2022}
I.~Adamovich, S.~Agarwal, E.~Ahedo, {\em et~al.}, ``The 2022 {{Plasma
  Roadmap}}: Low temperature plasma science and technology,'' {\em Journal of
  Physics D: Applied Physics} {\bf 55}(37), 373001  (2022).

\bibitem{anirudh2022ReviewDataDriven2023}
R.~Anirudh, R.~Archibald, M.~S. Asif, {\em et~al.}, ``2022 {{Review}} of
  {{Data-Driven Plasma Science}},'' {\em IEEE Transactions on Plasma Science}
  {\bf 51}, 1750--1838  (2023).

\bibitem{bonzaniniFoundationsMachineLearning2023}
A.~D. Bonzanini, K.~Shao, D.~B. Graves, {\em et~al.}, ``Foundations of machine
  learning for low-temperature plasmas: Methods and case studies,'' {\em Plasma
  Sources Science and Technology} {\bf 32}, 024003  (2023).

\bibitem{kambaraSciencebasedDatadrivenDevelopments2023}
M.~Kambara, S.~Kawaguchi, H.~J. Lee, {\em et~al.}, ``Science-based, data-driven
  developments in plasma processing for material synthesis and
  device-integration technologies,'' {\em Japanese Journal of Applied Physics}
  {\bf 62}(SA), SA0803  (2023).

\bibitem{shadmehrPrincipalComponentAnalysis1992}
R.~Shadmehr, D.~Angell, P.~B. Chou, {\em et~al.}, ``Principal {{Component
  Analysis}} of {{Optical Emission Spectroscopy}} and {{Mass Spectrometry}}:
  {{Application}} to {{Reactive Ion Etch Process Parameter Estimation Using
  Neural Networks}},'' {\em Journal of The Electrochemical Society} {\bf 139},
  907--914  (1992).

\bibitem{sangjeenhongNeuralNetworkModeling2003}
{Sang Jeen Hong}, G.~S. May, and {Dong-Cheol Park}, ``Neural network modeling
  of reactive ion etching using optical emission spectroscopy data,'' {\em IEEE
  Transactions on Semiconductor Manufacturing} {\bf 16}, 598--608  (2003).

\bibitem{himmelAdvantagesPlasmaEtch1993}
C.~Himmel and G.~May, ``Advantages of plasma etch modeling using neural
  networks over statistical techniques,'' {\em IEEE Transactions on
  Semiconductor Manufacturing} {\bf 6}, 103--111  (1993).

\bibitem{bleakieGrowingStructureMultiple2016}
A.~Bleakie and D.~Djurdjanovic, ``Growing {{Structure Multiple Model System}}
  for {{Quality Estimation}} in {{Manufacturing Processes}},'' {\em IEEE
  Transactions on Semiconductor Manufacturing} {\bf 29}, 79--97  (2016).

\bibitem{sangjeenhongNeuralnetworkbasedSensorFusion2005}
{Sang Jeen Hong} and G.~S. May, ``Neural-network-based sensor fusion of optical
  emission and mass spectroscopy data for real-time fault detection in reactive
  ion etching,'' {\em IEEE Transactions on Industrial Electronics} {\bf 52},
  1063--1072  (2005).

\bibitem{sutharNextgenerationVirtualMetrology2019}
K.~Suthar, D.~Shah, J.~Wang, {\em et~al.}, ``Next-generation virtual metrology
  for semiconductor manufacturing: {{A}} feature-based framework,'' {\em
  Computers \& Chemical Engineering} {\bf 127}, 140--149  (2019).

\bibitem{maggipintoDeepVMDeepLearningbased2019}
M.~Maggipinto, A.~Beghi, S.~McLoone, {\em et~al.}, ``{{DeepVM}}: {{A Deep
  Learning-based}} approach with automatic feature extraction for {{2D}} input
  data {{Virtual Metrology}},'' {\em Journal of Process Control} {\bf 84},
  24--34  (2019).

\bibitem{parkMicrorangeUniformityControl2021}
S.~Park, J.~Seong, Y.~Noh, {\em et~al.}, ``Micro-range uniformity control of
  the etching profile in the {{OLED}} display mass production referring to the
  {{PI-VM}} model,'' {\em Physics of Plasmas} {\bf 28}(10), 103505  (2021).

\bibitem{vandergaagArbitraryEEDFDetermination2021}
T.~{van der Gaag}, H.~Onishi, and H.~Akatsuka, ``Arbitrary {{EEDF}}
  determination of atmospheric-pressure plasma by applying machine learning to
  {{OES}} measurement,'' {\em Physics of Plasmas} {\bf 28}, 033511  (2021).

\bibitem{shaoActiveLearningguidedExploration2022}
K.~Shao, X.~Pei, D.~B. Graves, {\em et~al.}, ``Active learning-guided
  exploration of parameter space of air plasmas to enhance the energy
  efficiency of {{NO}}\${\textbackslash}less\$sub\${\textbackslash}greater\$ x
  \${\textbackslash}less\$/sub\${\textbackslash}greater\$ production,'' {\em
  Plasma Sources Science and Technology} {\bf 31}, 055018  (2022).

\bibitem{zhangSolvingPoissonEquation2019}
Z.~Zhang, L.~Zhang, Z.~Sun, {\em et~al.}, ``Solving {{Poisson}}'s {{Equation}}
  using {{Deep Learning}} in {{Particle Simulation}} of {{PN Junction}},'' in
  {\em 2019 {{Joint International Symposium}} on {{Electromagnetic
  Compatibility}}, {{Sapporo}} and {{Asia-Pacific International Symposium}} on
  {{Electromagnetic Compatibility}} ({{EMC Sapporo}}/{{APEMC}})},  305--308
  (2019).

\bibitem{trieschmannMachineLearningApproach2020a}
J.~Trieschmann, T.~Gergs, Y.~Liu, {\em et~al.}, ``A machine learning approach
  to the solution of {{Poisson}}'s equations for plasma simulations,'' in {\em
  Bulletin of the {{American Physical Society}}},  MW3.002, {APS}, ({Virtual
  Conference})  (2020).

\bibitem{dissanayakeNeuralnetworkbasedApproximationsSolving1994}
M.~W. M.~G. Dissanayake and N.~Phan-Thien, ``Neural-network-based
  approximations for solving partial differential equations,'' {\em
  Communications in Numerical Methods in Engineering} {\bf 10}(3), 195--201
  (1994).

\bibitem{aguilarDeepLearningBasedParticleinCell2021}
X.~Aguilar and S.~Markidis, ``A {{Deep Learning-Based Particle-in-Cell Method}}
  for {{Plasma Simulations}},'' in {\em 2021 {{IEEE International Conference}}
  on {{Cluster Computing}} ({{CLUSTER}})},  692--697  (2021).

\bibitem{rumelhartLearningInternalRepresentations1985}
D.~E. Rumelhart, G.~E. Hinton, and R.~J. Williams, ``Learning {{Internal
  Representations}} by {{Error Propagation}},'' {{ICS Report}} 8506,
  {University of California}, {San Diego (California), USA}  (1985).

\bibitem{fukushimaNeocognitronSelforganizingNeural1980}
K.~Fukushima, ``Neocognitron: {{A}} self-organizing neural network model for a
  mechanism of pattern recognition unaffected by shift in position,'' {\em
  Biological Cybernetics} {\bf 36}, 193--202  (1980).

\bibitem{chengUsingNeuralNetworks2021}
L.~Cheng, E.~A. Illarramendi, G.~Bogopolsky, {\em et~al.}, ``Using neural
  networks to solve the {{2D Poisson}} equation for electric field computation
  in plasma fluid simulations,''  (2021).

\bibitem{ozbayPoissonCNNConvolutional2021}
A.~G. {\"O}zbay, A.~Hamzehloo, S.~Laizet, {\em et~al.}, ``Poisson {{CNN}}:
  {{Convolutional}} neural networks for the solution of the {{Poisson}}
  equation on a {{Cartesian}} mesh,'' {\em Data-Centric Engineering} {\bf 2},
  e6  (2021).

\bibitem{ronnebergerUNetConvolutionalNetworks2015}
O.~Ronneberger, P.~Fischer, and T.~Brox, ``U-{{Net}}: {{Convolutional
  Networks}} for {{Biomedical Image Segmentation}},'' in {\em Medical {{Image
  Computing}} and {{Computer-Assisted Intervention}} {\textendash} {{MICCAI}}
  2015},  N.~Navab, J.~Hornegger, W.~M. Wells, {\em et~al.}, Eds., {\em Lecture
  {{Notes}} in {{Computer Science}}}, 234--241, {Springer International
  Publishing}, ({Cham})  (2015).

\bibitem{zengPhysicsInformedNeural2023}
X.~Zeng, S.~Zhang, C.~Ren, {\em et~al.}, ``Physics informed neural networks for
  electric field distribution characteristics analysis,'' {\em Journal of
  Physics D: Applied Physics} {\bf 56}, 165202  (2023).

\bibitem{raissiPhysicsinformedNeuralNetworks2019}
M.~Raissi, P.~Perdikaris, and G.~E. Karniadakis, ``Physics-informed neural
  networks: {{A}} deep learning framework for solving forward and inverse
  problems involving nonlinear partial differential equations,'' {\em Journal
  of Computational Physics} {\bf 378}, 686--707  (2019).

\bibitem{luDeepXDEDeepLearning2020}
L.~Lu, X.~Meng, Z.~Mao, {\em et~al.}, ``{{DeepXDE}}: {{A}} deep learning
  library for solving differential equations,'' {\em arXiv:1907.04502 [physics,
  stat]}   (2020).

\bibitem{baydinAutomaticDifferentiationMachine2018}
A.~G. Baydin, B.~A. Pearlmutter, A.~A. Radul, {\em et~al.}, ``Automatic
  {{Differentiation}} in {{Machine Learning}}: A {{Survey}},'' {\em Journal of
  Machine Learning Research} {\bf 18}, 43  (2018).

\bibitem{siffaMachineLearnedPoissonSolver2023}
I.~Siffa, M.~M. Becker, K.-D. Weltmann, {\em et~al.}, ``Towards a
  {{Machine-Learned Poisson Solver}} for {{Low-Temperature Plasma Simulations}}
  in {{Complex Geometries}},''  (2023).

\bibitem{huangDenselyConnectedConvolutional2017}
G.~Huang, Z.~Liu, L.~{van der Maaten}, {\em et~al.}, ``Densely {{Connected
  Convolutional Networks}},'' in {\em {{IEEE Conference}} on {{Computer
  Vision}} and {{Pattern Recognition}} ({{CVPR}})},  2261--2269, ({Honolulu,
  HI, USA})  (2017).

\bibitem{vaswaniAttentionAllYou2017}
A.~Vaswani, N.~Shazeer, N.~Parmar, {\em et~al.}, ``Attention is {{All}} you
  {{Need}},'' in {\em Advances in {{Neural Information Processing Systems}}},
  {\bf 30}, {Curran Associates, Inc.}  (2017).

\bibitem{chenTransUNetTransformersMake2021}
J.~Chen, Y.~Lu, Q.~Yu, {\em et~al.}, ``{{TransUNet}}: {{Transformers Make
  Strong Encoders}} for {{Medical Image Segmentation}},''  (2021).

\bibitem{wangImageQualityAssessment2004}
Z.~Wang, A.~Bovik, H.~Sheikh, {\em et~al.}, ``Image quality assessment: From
  error visibility to structural similarity,'' {\em IEEE Transactions on Image
  Processing} {\bf 13}, 600--612  (2004).

\bibitem{mihirdesaiDeepLearningArchitectureBasedApproach2022}
{Mihir Desai}, {Pratik Ghosh}, {Ahlad Kumar}, {\em et~al.}, ``Deep-{{Learning
  Architecture-Based Approach}} for 2-{{D-Simulation}} of {{Microwave Plasma
  Interaction}},'' {\em IEEE Transactions on Microwave Theory and Techniques}
  {\bf 70}(12), 5359--5368  (2022).

\bibitem{rahamanSpectralBiasNeural2019}
N.~Rahaman, A.~Baratin, D.~Arpit, {\em et~al.}, ``On the {{Spectral Bias}} of
  {{Neural Networks}},'' in {\em Proceedings of the 36th {{International
  Conference}} on {{Machine Learning}}},  5301--5310, {PMLR}  (2019).

\bibitem{vermaSurrogateModelsLow2021}
A.~K. Verma, X.~Li, S.~Ganta, {\em et~al.}, ``Surrogate {{Models}} for {{Low
  Temperature Plasma Simulations}} with {{Deep Learning}},'' in {\em Bulletin
  of the {{American Physical Society}}},  PR24.003, {APS}, ({Virtual
  Conference})  (2021).

\bibitem{ichikawaConstructionSurrogateModel2022}
M.~Ichikawa, K.~Ikuse, K.-L. Chen, {\em et~al.}, ``Construction of a surrogate
  model for low-temperature plasma simulation using machine learning,'' in {\em
  69th {{Japan Society}} of {{Applied Physics}} ({{JSAP}}) {{Spring Meeting}}
  2022},   (2022).

\bibitem{kushnerHybridModellingLow2009}
M.~J. Kushner, ``Hybrid modelling of low temperature plasmas for fundamental
  investigations and equipment design,'' {\em Journal of Physics D: Applied
  Physics} {\bf 42}(19), 194013  (2009).

\bibitem{koComputationalApproachPlasma2023}
J.~Ko, J.~Bae, M.~Park, {\em et~al.}, ``Computational approach for plasma
  process optimization combined with deep learning model,'' {\em Journal of
  Physics D: Applied Physics} {\bf 56}, 344001  (2023).

\bibitem{zhangEfficientNumericalSimulation2023a}
Y.-T. Zhang, S.-H. Gao, and F.~Ai, ``Efficient numerical simulation of
  atmospheric pulsed discharges by introducing deep learning,'' {\em Frontiers
  in Physics} {\bf 11}  (2023).

\bibitem{zhangEfficientNumericalSimulation2023}
Y.-T. Zhang, S.-H. Gao, and Y.-Y. Zhu, ``Efficient numerical simulation on
  dielectric barrier discharges at atmospheric pressure integrated by deep
  neural network,'' {\em Journal of Applied Physics} {\bf 133}, 053303  (2023).

\bibitem{wangModelingDischargeCharacteristics2023}
X.-C. Wang and Y.-T. Zhang, ``Modeling of discharge characteristics and plasma
  chemistry in atmospheric {{CO2}} pulsed plasmas employing deep neural
  network,'' {\em Journal of Applied Physics} {\bf 133}, 143301  (2023).

\bibitem{kawaguchiDeepLearningSolving2020}
S.~Kawaguchi, K.~Takahashi, H.~Ohkama, {\em et~al.}, ``Deep learning for
  solving the {{Boltzmann}} equation of electrons in weakly ionized plasma,''
  {\em Plasma Sources Science and Technology} {\bf 29}, 025021  (2020).

\bibitem{kawaguchiPhysicsinformedNeuralNetworks2022}
S.~Kawaguchi and T.~Murakami, ``Physics-informed neural networks for solving
  the {{Boltzmann}} equation of the electron velocity distribution function in
  weakly ionized plasmas,'' {\em Japanese Journal of Applied Physics} {\bf 61},
  086002  (2022).

\bibitem{wangUnderstandingMitigatingGradient2021}
S.~Wang, Y.~Teng, and P.~Perdikaris, ``Understanding and {{Mitigating Gradient
  Flow Pathologies}} in {{Physics-Informed Neural Networks}},'' {\em SIAM
  Journal on Scientific Computing} {\bf 43}, A3055--A3081  (2021).

\bibitem{kimNumericalStrategySolving2023}
J.~S. Kim, K.~Denpoh, S.~Kawaguchi, {\em et~al.}, ``Numerical strategy for
  solving the {{Boltzmann}} equation with variable {{E}}/{{N}} using
  physics-informed neural networks,'' {\em Journal of Physics D: Applied
  Physics} {\bf 56}, 344002  (2023).

\bibitem{zhongDeepLearningThermal2020}
L.~Zhong, Q.~Gu, and B.~Wu, ``Deep learning for thermal plasma simulation:
  {{Solving}} 1-{{D}} arc model as an example,'' {\em Computer Physics
  Communications} {\bf 257}, 107496  (2020).

\bibitem{zhongLowtemperaturePlasmaSimulation2022}
L.~Zhong, B.~Wu, and Y.~Wang, ``Low-temperature plasma simulation based on
  physics-informed neural networks: {{Frameworks}} and preliminary
  applications,'' {\em Physics of Fluids} {\bf 34}, 087116  (2022).

\bibitem{zhongAcceleratingPhysicsinformedNeural2023}
L.~Zhong, B.~Wu, and Y.~Wang, ``Accelerating physics-informed neural network
  based {{1D}} arc simulation by meta learning,'' {\em Journal of Physics D:
  Applied Physics} {\bf 56}, 074006  (2023).

\bibitem{psarosMetalearningPINNLoss2022}
A.~F. Psaros, K.~Kawaguchi, and G.~E. Karniadakis, ``Meta-learning {{PINN}}
  loss functions,'' {\em Journal of Computational Physics} {\bf 458}, 111121
  (2022).

\bibitem{kawaguchiDatadrivenDiscoveryElectron2023}
S.~Kawaguchi, K.~Takahashi, and K.~Satoh, ``Data-driven discovery of electron
  continuity equations in electron swarm map for determining electron transport
  coefficients in argon,'' {\em Journal of Physics D: Applied Physics} {\bf
  56}, 244003  (2023).

\bibitem{xiaoMultiscaleModelingRecurrent2021}
T.~Xiao and D.~Ni, ``Multiscale {{Modeling}} and {{Recurrent Neural Network
  Based Optimization}} of a {{Plasma Etch Process}},'' {\em Processes} {\bf 9},
  151  (2021).

\bibitem{xiaoRecurrentNeuralNetworkBasedModel2022}
T.~Xiao, Z.~Wu, P.~D. Christofides, {\em et~al.}, ``Recurrent
  {{Neural-Network-Based Model Predictive Control}} of a {{Plasma Etch
  Process}},'' {\em Industrial \& Engineering Chemistry Research} {\bf 61},
  638--652  (2022).

\bibitem{gravesReportScienceChallenges2023}
``Report on {{Science Challenges}} and {{Research Opportunities}} for {{Plasma
  Applications}} in {{Microelectronics}},'' tech. rep., {Department of Energy
  Office of Science Fusion Energy Sciences Workshop}  (2023).

\bibitem{turnerUncertaintySensitivityAnalysis2016}
M.~M. Turner, ``Uncertainty and sensitivity analysis in complex plasma
  chemistry models,'' {\em Plasma Sources Science and Technology} {\bf 25}(1),
  015003  (2016).

\bibitem{alvesFoundationsPlasmaStandards2023}
L.~L. Alves, M.~M. Becker, J.~van Dijk, {\em et~al.}, ``Foundations of plasma
  standards,'' {\em Plasma Sources Science and Technology} {\bf 32}, 023001
  (2023).

\bibitem{morganUseNumericalOptimization1991}
W.~L. Morgan, ``Use of numerical optimization algorithms to obtain cross
  sections from electron swarm data,'' {\em Physical Review A} {\bf 44},
  1677--1681  (1991).

\bibitem{stokesDeterminingCrossSections2020}
P.~W. Stokes, D.~G. Cocks, M.~J. Brunger, {\em et~al.}, ``Determining cross
  sections from transport coefficients using deep neural networks,'' {\em
  Plasma Sources Science and Technology} {\bf 29}, 055009  (2020).

\bibitem{stokesImprovedSetElectronTHFA2021}
P.~W. Stokes, S.~P. Foster, M.~J.~E. Casey, {\em et~al.}, ``An improved set of
  electron-{{THFA}} cross sections refined through a neural network-based
  analysis of swarm data,'' {\em The Journal of Chemical Physics} {\bf 154},
  084306  (2021).

\bibitem{lxcat-teamHttpsNlLxcat2023}
{LXCat-Team}, ``{{https://nl.lxcat.net/home/}},''  (2023).

\bibitem{jetlyExtractingElectronScattering2021}
V.~Jetly and B.~Chaudhury, ``Extracting electron scattering cross sections from
  swarm data using deep neural networks,'' {\em Machine Learning: Science and
  Technology} {\bf 2}, 035025  (2021).

\bibitem{hagelaarSolvingBoltzmannEquation2005}
G.~J.~M. Hagelaar and L.~C. Pitchford, ``Solving the {{Boltzmann}} equation to
  obtain electron transport coefficients and rate coefficients for fluid
  models,'' {\em Plasma Sources Science and Technology} {\bf 14}, 722  (2005).

\bibitem{zhongFastPredictionElectronimpact2019}
L.~Zhong, ``Fast prediction of electron-impact ionization cross sections of
  large molecules via machine learning,'' {\em Journal of Applied Physics} {\bf
  125}, 183302  (2019).

\bibitem{harrisDataDrivenMachineLearning2023}
A.~L. Harris and J.~Nepomuceno, ``A {{Data-Driven Machine Learning Approach}}
  for {{Electron-Molecule Ionization Cross Sections}},''  (2023).

\bibitem{reiserDeterminingChemicalReaction2021}
D.~Reiser, A.~{von Keudell}, and T.~Urbanietz, ``Determining {{Chemical
  Reaction Systems}} in {{Plasma-Assisted Conversion}} of {{Methane Using
  Genetic Algorithms}},'' {\em Plasma Chemistry and Plasma Processing} {\bf
  41}, 793--813  (2021).

\bibitem{hanicinecRegressionModelPlasma2023}
M.~Hanicinec, S.~Mohr, and J.~Tennyson, ``A regression model for plasma
  reaction kinetics,'' {\em Journal of Physics D: Applied Physics} {\bf 56},
  374001  (2023).

\bibitem{tennysonQDBNewDatabase2017}
J.~Tennyson, S.~Rahimi, C.~Hill, {\em et~al.}, ``{{QDB}}: A new database of
  plasma chemistries and reactions,'' {\em Plasma Sources Science and
  Technology} {\bf 26}(5), 055014  (2017).

\bibitem{wakelamKineticDatabaseAstrochemistry2012}
V.~Wakelam, E.~Herbst, J.-C. Loison, {\em et~al.}, ``A {{Kinetic Database}} for
  {{Astrochemistry}} ({{KIDA}}),'' {\em The Astrophysical Journal Supplement
  Series} {\bf 199}, 21  (2012).

\bibitem{parkNewVersionPlasma2020}
J.-H. Park, H.~Choi, W.-S. Chang, {\em et~al.}, ``A {{New Version}} of the
  {{Plasma Database}} for {{Plasma Physics}} in the {{Data Center}} for
  {{Plasma Properties}},'' {\em Applied Science and Convergence Technology}
  {\bf 29}, 5--9  (2020).

\bibitem{mcelroyUMISTDatabaseAstrochemistry2013}
D.~McElroy, C.~Walsh, A.~J. Markwick, {\em et~al.}, ``The {{UMIST}} database
  for astrochemistry 2012,'' {\em Astronomy \& Astrophysics} {\bf 550}, A36
  (2013).

\bibitem{jolliffePrincipalComponentAnalysis2016}
I.~T. Jolliffe and J.~Cadima, ``Principal component analysis: A review and
  recent developments,'' {\em Philosophical Transactions of the Royal Society
  A: Mathematical, Physical and Engineering Sciences} {\bf 374}, 20150202
  (2016).

\bibitem{maasSimplifyingChemicalKinetics1992}
U.~Maas and S.~B. Pope, ``Simplifying chemical kinetics: {{Intrinsic}}
  low-dimensional manifolds in composition space,'' {\em Combustion and Flame}
  {\bf 88}, 239--264  (1992).

\bibitem{peerenboomDimensionReductionNonequilibrium2015}
K.~Peerenboom, A.~Parente, T.~Koz{\'a}k, {\em et~al.}, ``Dimension reduction of
  non-equilibrium plasma kinetic models using principal component analysis,''
  {\em Plasma Sources Science and Technology} {\bf 24}, 025004  (2015).

\bibitem{rehmanSimplifyingPlasmaChemistry2016}
T.~Rehman, E.~Kemaneci, W.~Graef, {\em et~al.}, ``Simplifying plasma chemistry
  via {{ILDM}},'' {\em Journal of Physics: Conference Series} {\bf 682}, 012035
   (2016).

\bibitem{sakaiAnalysisWeblikeNetwork2015}
O.~Sakai, K.~Nobuto, S.~Miyagi, {\em et~al.}, ``Analysis of weblike network
  structures of directed graphs for chemical reactions in methane plasmas,''
  {\em AIP Advances} {\bf 5}, 107140  (2015).

\bibitem{sakaiComplexityVisualizationDataset2022}
O.~Sakai, S.~Kawaguchi, and T.~Murakami, ``Complexity visualization, dataset
  acquisition, and machine-learning perspectives for low-temperature plasma: A
  review,'' {\em Japanese Journal of Applied Physics} {\bf 61}, 070101  (2022).

\bibitem{sunChemistryReductionComplex2020}
S.~R. Sun, H.~X. Wang, and A.~Bogaerts, ``Chemistry reduction of complex
  {{CO2}} chemical kinetics: Application to a gliding arc plasma,'' {\em Plasma
  Sources Science and Technology} {\bf 29}, 025012  (2020).

\bibitem{murakamiRescalingComplexNetwork2020}
T.~Murakami and O.~Sakai, ``Rescaling the complex network of low-temperature
  plasma chemistry through graph-theoretical analysis,'' {\em Plasma Sources
  Science and Technology} {\bf 29}, 115018  (2020).

\bibitem{mizuiGraphicalClassificationMultiCentralityIndex2017}
Y.~Mizui, T.~Kojima, S.~Miyagi, {\em et~al.}, ``Graphical {{Classification}} in
  {{Multi-Centrality-Index Diagrams}} for {{Complex Chemical Networks}},'' {\em
  Symmetry} {\bf 9}, 309  (2017).

\bibitem{holmesGraphTheoryApplied2021}
T.~D. Holmes, R.~H. Rothman, and W.~B. Zimmerman, ``Graph {{Theory Applied}} to
  {{Plasma Chemical Reaction Engineering}},'' {\em Plasma Chemistry and Plasma
  Processing} {\bf 41}, 531--557  (2021).

\bibitem{venturiUncertaintyawareStrategyPlasma2023}
S.~Venturi, W.~Yang, I.~Kaganovich, {\em et~al.}, ``An uncertainty-aware
  strategy for plasma mechanism reduction with directed weighted graphs,'' {\em
  Physics of Plasmas} {\bf 30}, 043904  (2023).

\bibitem{zhuTailoringElectricField2022}
Y.~Zhu, Y.~Bo, X.~Chen, {\em et~al.}, ``Tailoring electric field signals of
  nonequilibrium discharges by the deep learning method and physical
  corrections,'' {\em Plasma Processes and Polymers} {\bf 19}(3), e2100155
  (2022).

\bibitem{panDeepLearningassistedPulsed2023}
J.~Pan, Y.~Liu, S.~Zhang, {\em et~al.}, ``Deep learning-assisted pulsed
  discharge plasma catalysis modeling,'' {\em Energy Conversion and Management}
  {\bf 277}, 116620  (2023).

\bibitem{biersackMonteCarloComputer1980}
J.~P. Biersack and L.~G. Haggmark, ``A {{Monte Carlo}} computer program for the
  transport of energetic ions in amorphous targets,'' {\em Nuclear Instruments
  and Methods} {\bf 174}, 257  (1980).

\bibitem{dingMicroscopicModelingOptimal2020}
Y.~Ding, Y.~Zhang, G.~Orkoulas, {\em et~al.}, ``Microscopic modeling and
  optimal operation of plasma enhanced atomic layer deposition,'' {\em Chemical
  Engineering Research and Design} {\bf 159}, 439--454  (2020).

\bibitem{burdenBayesianRegularizationNeural2009}
F.~Burden and D.~Winkler, ``Bayesian {{Regularization}} of {{Neural
  Networks}},'' in {\em Artificial {{Neural Networks}}: {{Methods}} and
  {{Applications}}},  D.~J. Livingstone, Ed., {\em Methods in {{Molecular
  Biology}}}, 23--42, {Humana Press}, {Totowa, NJ}  (2009).

\bibitem{dingMachineLearningbasedModeling2021}
Y.~Ding, Y.~Zhang, H.~Y. Chung, {\em et~al.}, ``Machine learning-based modeling
  and operation of plasma-enhanced atomic layer deposition of hafnium oxide
  thin films,'' {\em Computers \& Chemical Engineering} {\bf 144}, 107148
  (2021).

\bibitem{sheilPreciseControlNanoscale2021}
R.~Sheil, J.~M.~P. Martirez, X.~Sang, {\em et~al.}, ``Precise {{Control}} of
  {{Nanoscale Cu Etching}} via {{Gas-Phase Oxidation}} and {{Chemical
  Complexation}},'' {\em The Journal of Physical Chemistry C} {\bf 125},
  1819--1832  (2021).

\bibitem{xiaPlasmaOxidationCopper2022}
Y.~Xia and P.~Sautet, ``Plasma {{Oxidation}} of {{Copper}}: {{Molecular
  Dynamics Study}} with {{Neural Network Potentials}},'' {\em ACS Nano} {\bf
  16}, 20680--20692  (2022).

\bibitem{aktulgaParallelReactiveMolecular2012}
H.~M. Aktulga, J.~C. Fogarty, S.~A. Pandit, {\em et~al.}, ``Parallel reactive
  molecular dynamics: {{Numerical}} methods and algorithmic techniques,'' {\em
  Parallel Computing} {\bf 38}, 245--259  (2012).

\bibitem{zhuDevelopmentReactiveForce2020}
W.~Zhu, H.~Gong, Y.~Han, {\em et~al.}, ``Development of a {{Reactive Force
  Field}} for {{Simulations}} on the {{Catalytic Conversion}} of
  {{C}}/{{H}}/{{O Molecules}} on {{Cu-Metal}} and {{Cu-Oxide Surfaces}} and
  {{Application}} to {{Cu}}/{{CuO-Based Chemical Looping}},'' {\em The Journal
  of Physical Chemistry C} {\bf 124}, 12512--12520  (2020).

\bibitem{wanDeepLearningAssistedInvestigation2022a}
M.~Wan, H.~Yue, J.~Notarangelo, {\em et~al.}, ``Deep {{Learning-Assisted
  Investigation}} of {{Electric Field}}{\textendash}{{Dipole Effects}} on
  {{Catalytic Ammonia Synthesis}},'' {\em JACS Au} {\bf 2}, 1338--1349  (2022).

\bibitem{scarselliGraphNeuralNetwork2009}
F.~Scarselli, M.~Gori, A.~C. Tsoi, {\em et~al.}, ``The {{Graph Neural Network
  Model}},'' {\em IEEE Transactions on Neural Networks} {\bf 20}, 61--80
  (2009).

\bibitem{wuComprehensiveSurveyGraph2021}
Z.~Wu, S.~Pan, F.~Chen, {\em et~al.}, ``A {{Comprehensive Survey}} on {{Graph
  Neural Networks}},'' {\em IEEE Transactions on Neural Networks and Learning
  Systems} {\bf 32}, 4--24  (2021).

\bibitem{kipfSemiSupervisedClassificationGraph2022}
T.~N. Kipf and M.~Welling, ``Semi-{{Supervised Classification}} with {{Graph
  Convolutional Networks}},'' in {\em International {{Conference}} on
  {{Learning Representations}}},   (2022).

\bibitem{kedaloApplicabilityFridmanMacheret2023}
Y.~M. Kedalo, A.~A. Knizhnik, and B.~V. Potapkin, ``Applicability of the
  {{Fridman}}{\textendash}{{Macheret}} {$\alpha$}-{{Model}} to {{Heterogeneous
  Processes}} in the {{Case}} of {{Dissociative Adsorption}} of {{N2}} on the
  {{Ru Surface}},'' {\em The Journal of Physical Chemistry C} {\bf 127},
  11536--11541  (2023).

\bibitem{fridmanPlasmaChemistry2008}
A.~Fridman, {\em Plasma {{Chemistry}}}, {Cambridge University Press}, {New
  York, NY, USA}  (2008).

\bibitem{bartokRepresentingChemicalEnvironments2013}
A.~P. Bart{\'o}k, R.~Kondor, and G.~Cs{\'a}nyi, ``On representing chemical
  environments,'' {\em Physical Review B} {\bf 87}, 184115  (2013).

\bibitem{balQuantifyingImpactVibrational2021}
K.~M. Bal and E.~C. Neyts, ``Quantifying the impact of vibrational
  nonequilibrium in plasma catalysis: Insights from a molecular dynamics model
  of dissociative chemisorption,'' {\em Journal of Physics D: Applied Physics}
  {\bf 54}, 394004  (2021).

\bibitem{mendelsCollectiveVariablesLocal2018}
D.~Mendels, G.~Piccini, and M.~Parrinello, ``Collective {{Variables}} from
  {{Local Fluctuations}},'' {\em The Journal of Physical Chemistry Letters}
  {\bf 9}, 2776--2781  (2018).

\bibitem{valssonVariationalApproachEnhanced2014}
O.~Valsson and M.~Parrinello, ``Variational {{Approach}} to {{Enhanced
  Sampling}} and {{Free Energy Calculations}},'' {\em Physical Review Letters}
  {\bf 113}, 090601  (2014).

\bibitem{barducciWellTemperedMetadynamicsSmoothly2008}
A.~Barducci, G.~Bussi, and M.~Parrinello, ``Well-{{Tempered Metadynamics}}: {{A
  Smoothly Converging}} and {{Tunable Free-Energy Method}},'' {\em Physical
  Review Letters} {\bf 100}, 020603  (2008).

\bibitem{laioEscapingFreeenergyMinima2002}
A.~Laio and M.~Parrinello, ``Escaping free-energy minima,'' {\em Proceedings of
  the National Academy of Sciences} {\bf 99}, 12562--12566  (2002).

\bibitem{sironEnablingAutomatedHighthroughput2023}
M.~Siron, N.~Chandrasekhar, and K.~A. Persson, ``Enabling automated
  high-throughput {{Density Functional Theory}} studies of amorphous material
  surface reactions,'' {\em Computational Materials Science} {\bf 226}, 112192
  (2023).

\bibitem{hamedaniPrimaryRadiationDamage2021}
A.~Hamedani, J.~Byggm{\"a}star, F.~Djurabekova, {\em et~al.}, ``Primary
  radiation damage in silicon from the viewpoint of a machine learning
  interatomic potential,'' {\em Physical Review Materials} {\bf 5}, 114603
  (2021).

\bibitem{bartokGaussianApproximationPotentials2010}
A.~P. Bart{\'o}k, M.~C. Payne, R.~Kondor, {\em et~al.}, ``Gaussian
  {{Approximation Potentials}}: {{The Accuracy}} of {{Quantum Mechanics}},
  without the {{Electrons}},'' {\em Physical Review Letters} {\bf 104}, 136403
  (2010).

\bibitem{nordlundRepulsiveInteratomicPotentials1997}
K.~Nordlund, N.~Runeberg, and D.~Sundholm, ``Repulsive interatomic potentials
  calculated using {{Hartree-Fock}} and density-functional theory methods,''
  {\em Nuclear Instruments and Methods in Physics Research Section B: Beam
  Interactions with Materials and Atoms} {\bf 132}, 45--54  (1997).

\bibitem{stillingerComputerSimulationLocal1985}
F.~H. Stillinger and T.~A. Weber, ``Computer simulation of local order in
  condensed phases of silicon,'' {\em Physical Review B} {\bf 31}, 5262--5271
  (1985).

\bibitem{tersoffNewEmpiricalApproach1988}
J.~Tersoff, ``New empirical approach for the structure and energy of covalent
  systems,'' {\em Physical Review B} {\bf 37}, 6991--7000  (1988).

\bibitem{hamedaniInsightsPrimaryRadiation2020}
A.~Hamedani, J.~Byggm{\"a}star, F.~Djurabekova, {\em et~al.}, ``Insights into
  the primary radiation damage of silicon by a machine learning interatomic
  potential,'' {\em Materials Research Letters} {\bf 8}, 364--372  (2020).

\bibitem{yamamuraEnergyDependenceIonInduced1996}
Y.~Yamamura and H.~Tawara, ``Energy {{Dependence}} of {{Ion-Induced Sputtering
  Yields}} from {{Monatomic Solids}} at {{Normal Incidence}},'' {\em Atomic
  Data and Nuclear Data Tables} {\bf 62}(2), 149  (1996).

\bibitem{phadkeSputterYieldsMonoatomic2022}
P.~Phadke, A.~A. Zameshin, J.~M. Sturm, {\em et~al.}, ``Sputter yields of
  monoatomic solids by {{Ar}} and {{Ne}} ions near the threshold: {{A
  Bayesian}} analysis of the {{Yamamura Model}},'' {\em Nuclear Instruments and
  Methods in Physics Research Section B: Beam Interactions with Materials and
  Atoms} {\bf 520}, 29--39  (2022).

\bibitem{mollerTridynTRIMSimulation1984}
W.~M{\"o}ller and W.~Eckstein, ``Tridyn - {{A TRIM}} simulation code including
  dynamic composition changes,'' {\em Nuclear Instruments and Methods in
  Physics Research Section B} {\bf 2}, 814  (1984).

\bibitem{ecksteinNewFitFormulae2003}
W.~Eckstein and R.~Preuss, ``New fit formulae for the sputtering yield,'' {\em
  Journal of Nuclear Materials} {\bf 320}, 209  (2003).

\bibitem{preussBayesianDeterminationParameters2019}
R.~Preuss, R.~Arredondo, and U.~{von Toussaint}, ``Bayesian {{Determination}}
  of {{Parameters}} for {{Plasma-Wall Interactions}},'' {\em Entropy} {\bf
  21}(12), 1175  (2019).

\bibitem{ecksteinSDTrimSPMonteCarloCode2007}
W.~Eckstein, R.~Dohmen, A.~Mutzke, {\em et~al.}, ``{{SDTrimSP}}: {{Ein
  Monte-Carlo Code}} zur {{Berechnung}} von {{Stossereignissen}} in
  ungeordneten {{Targets}},'' Tech. Rep. 12/3, {Max-Planck-Institut f{\"u}r
  Plasmaphysik}, {Garching}  (2007).

\bibitem{wienerHomogeneousChaos1938}
N.~Wiener, ``The {{Homogeneous Chaos}},'' {\em American Journal of Mathematics}
  {\bf 60}(4), 897--936  (1938).

\bibitem{krugerMachineLearningPlasmasurface2019}
F.~Kr{\"u}ger, T.~Gergs, and J.~Trieschmann, ``Machine learning plasma-surface
  interface for coupling sputtering and gas-phase transport simulations,'' {\em
  Plasma Sources Science and Technology} {\bf 28}(3), 035002  (2019).

\bibitem{gergsEfficientPlasmasurfaceInteraction2022}
T.~Gergs, B.~Borislavov, and J.~Trieschmann, ``Efficient plasma-surface
  interaction surrogate model for sputtering processes based on autoencoder
  neural networks,'' {\em Journal of Vacuum Science \& Technology B} {\bf
  40}(1), 012802  (2022).

\bibitem{kingmaAutoEncodingVariationalBayes2013}
D.~P. Kingma and M.~Welling, ``Auto-{{Encoding Variational Bayes}},'' in {\em
  International {{Conference}} on {{Learning Representations}} 2014},   (2013).

\bibitem{rezendeStochasticBackpropagationApproximate2014}
D.~J. Rezende, S.~Mohamed, and D.~Wierstra, ``Stochastic {{Backpropagation}}
  and {{Approximate Inference}} in {{Deep Generative Models}},'' in {\em
  Proceedings of {{Machine Learning Research}}},   {\bf 32}, 1278--1286
  (2014).

\bibitem{higginsVVAELearningBasic2017}
I.~Higgins, L.~Matthey, A.~Pal, {\em et~al.}, ``{$\beta$}-{{VAE}}: {{Learning
  Basic Visual Concepts}} with a {{Constrained Variational Framework}},'' in
  {\em International {{Conference}} on {{Learning Representations}}},  22
  (2017).

\bibitem{kimDeepNeuralNetworkbased2023}
B.~Kim, J.~Bae, H.~Jeong, {\em et~al.}, ``Deep neural network-based
  reduced-order modeling of ion{\textendash}surface interactions combined with
  molecular dynamics simulation,'' {\em Journal of Physics D: Applied Physics}
  {\bf 56}, 384005  (2023).

\bibitem{vincentExtractingComposingRobust2008}
P.~Vincent, H.~Larochelle, Y.~Bengio, {\em et~al.}, ``Extracting and composing
  robust features with denoising autoencoders,'' in {\em Proceedings of the
  25th International Conference on Machine Learning},  1096--1103, {ACM Press},
  ({Helsinki, Finland})  (2008).

\bibitem{sohnLearningStructuredOutput2015}
K.~Sohn, H.~Lee, and X.~Yan, ``Learning {{Structured Output Representation}}
  using {{Deep Conditional Generative Models}},'' in {\em Advances in {{Neural
  Information Processing Systems}} 28},   (2015).

\bibitem{gergsPhysicsseparatingArtificialNeural2023}
T.~Gergs, T.~Mussenbrock, and J.~Trieschmann, ``Physics-separating artificial
  neural networks for predicting sputtering and thin film deposition of {{AlN}}
  in {{Ar}}/{{N2}} discharges on experimental timescales,'' {\em Journal of
  Physics D: Applied Physics} {\bf 56}(19), 194001  (2023).

\bibitem{balTimeScaleAssociated2014}
K.~M. Bal and E.~C. Neyts, ``On the time scale associated with {{Monte Carlo}}
  simulations,'' {\em The Journal of Chemical Physics} {\bf 141}, 204104
  (2014).

\bibitem{meesUniformacceptanceForcebiasMonte2012}
M.~J. Mees, G.~Pourtois, E.~C. Neyts, {\em et~al.}, ``Uniform-acceptance
  force-bias {{Monte Carlo}} method with time scale to study solid-state
  diffusion,'' {\em Physical Review B} {\bf 85}, 134301  (2012).

\bibitem{neytsCombiningMolecularDynamics2013}
E.~C. Neyts and A.~Bogaerts, ``Combining molecular dynamics with {{Monte
  Carlo}} simulations: Implementations and applications,'' {\em Theoretical
  Chemistry Accounts} {\bf 132}(2), 1320  (2013).

\bibitem{gergsMolecularDynamicsStudy2022}
T.~Gergs, T.~Mussenbrock, and J.~Trieschmann, ``Molecular dynamics study on the
  role of {{Ar}} ions in the sputter deposition of {{Al}} thin films,'' {\em
  Journal of Applied Physics} {\bf 132}(6), 063302  (2022).

\bibitem{gergsPhysicsseparatingArtificialNeural2023a}
T.~Gergs, T.~Mussenbrock, and J.~Trieschmann, ``Physics-separating artificial
  neural networks for predicting initial stages of {{Al}} sputtering and thin
  film deposition in {{Ar}} plasma discharges,'' {\em Journal of Physics D:
  Applied Physics} {\bf 56}(8), 084003  (2023).

\bibitem{jaramillo-boteroGeneralMultiobjectiveForce2014}
A.~{Jaramillo-Botero}, S.~Naserifar, and W.~A.~I. Goddard, ``General
  {{Multiobjective Force Field Optimization Framework}}, with {{Application}}
  to {{Reactive Force Fields}} for {{Silicon Carbide}},'' {\em Journal of
  Chemical Theory and Computation} {\bf 10}, 1426--1439  (2014).

\bibitem{gergsChargeoptimizedManybodyInteraction2023}
T.~Gergs, T.~Mussenbrock, and J.~Trieschmann, ``Charge-optimized many-body
  interaction potential for {{AlN}} revisited to explore
  plasma{\textendash}surface interactions,'' {\em Scientific Reports} {\bf
  13}(1), 5287  (2023).

\bibitem{bergModelingReactiveSputtering1987}
S.~Berg, H.-O. Blom, T.~Larsson, {\em et~al.}, ``Modeling of reactive
  sputtering of compound materials,'' {\em Journal of Vacuum Science \&
  Technology A} {\bf 5}, 202  (1987).

\bibitem{woelfelModelReductionIdentification2017}
C.~Woelfel, P.~Awakowicz, and J.~Lunze, ``Model reduction and identification of
  nonlinear reactive sputter processes,'' {\em IFAC-PapersOnLine} {\bf 50},
  13728--13734  (2017).

\bibitem{woelfelModelApproximationStabilization2019}
C.~Woelfel, D.~Bockhorn, P.~Awakowicz, {\em et~al.}, ``Model approximation and
  stabilization of reactive sputter processes,'' {\em Journal of Process
  Control} {\bf 83}, 121--128  (2019).

\bibitem{woelfelPlasmaStateControl2019}
C.~Woelfel, M.~Oberberg, P.~Awakowicz, {\em et~al.}, ``Plasma {{State Control}}
  of {{Reactive Sputter Processes}},'' in {\em Proceedings of the
  {{6thInternational Conference}} of {{Control}}, {{Dynamic Systems}}, and
  {{Robotics}} ({{CDSR}}'19)},  107, ({Ottawa, Canada})  (2019).

\bibitem{woelfelControlorientedPlasmaModeling2021}
C.~Woelfel, M.~Oberberg, B.~Berger, {\em et~al.}, ``Control-oriented plasma
  modeling and controller design for reactive sputtering,'' {\em IFAC Journal
  of Systems and Control} {\bf 16}, 100142  (2021).

\bibitem{gidonPredictiveControl2D2019}
D.~Gidon, D.~B. Graves, and A.~Mesbah, ``Predictive control of {{2D}} spatial
  thermal dose delivery in atmospheric pressure plasma jets,'' {\em Plasma
  Sources Science and Technology} {\bf 28}(8), 085001  (2019).

\bibitem{gidonDatadrivenLPVModel2021}
D.~Gidon, H.~S. Abbas, A.~D. Bonzanini, {\em et~al.}, ``Data-driven {{LPV}}
  model predictive control of a cold atmospheric plasma jet for biomaterials
  processing,'' {\em Control Engineering Practice} {\bf 109}, 104725  (2021).

\bibitem{rodriguesDataDrivenAdaptiveOptimal2023}
D.~Rodrigues, K.~J. Chan, and A.~Mesbah, ``Data-{{Driven Adaptive Optimal
  Control Under Model Uncertainty}}: {{An Application}} to {{Cold Atmospheric
  Plasmas}},'' {\em IEEE Transactions on Control Systems Technology} {\bf 31},
  55--69  (2023).

\bibitem{berkoozProperOrthogonalDecomposition1993}
G.~Berkooz, P.~Holmes, and J.~L. Lumley, ``The {{Proper Orthogonal
  Decomposition}} in the {{Analysis}} of {{Turbulent Flows}},'' {\em Annual
  Review of Fluid Mechanics} {\bf 25}(1), 539--575  (1993).

\bibitem{greveRealtimeStateEstimation2021}
C.~M. Greve, M.~Majji, and K.~Hara, ``Real-time state estimation of
  low-frequency plasma oscillations in {{Hall}} effect thrusters,'' {\em
  Physics of Plasmas} {\bf 28}, 093509  (2021).

\bibitem{greveEstimationPlasmaProperties2022}
C.~M. Greve and K.~Hara, ``Estimation of plasma properties using an extended
  {{Kalman}} filter with plasma global models,'' {\em Journal of Physics D:
  Applied Physics} {\bf 55}, 255201  (2022).

\bibitem{kanarikHumanMachineCollaboration2023}
K.~J. Kanarik, W.~T. Osowiecki, Y.~J. Lu, {\em et~al.},
  ``Human{\textendash}machine collaboration for improving semiconductor process
  development,'' {\em Nature} {\bf 616}, 707--711  (2023).

\bibitem{jaegerTutorialTrainingRecurrent2002}
H.~Jaeger, ``A tutorial on training recurrent neural networks, covering
  {{BPPT}}, {{RTRL}}, {{EKF}} and the "echo state network" approach,'' tech.
  rep., {GMD-Forschungszentrum Informationstechnik Bonn}, {Bonn}  (2002).

\bibitem{hochreiterLongShortTermMemory1997}
S.~Hochreiter and J.~Schmidhuber, ``Long {{Short-Term Memory}},'' {\em Neural
  Computation} {\bf 9}, 1735--1780  (1997).

\bibitem{gersLearningForgetContinual2000}
F.~A. Gers, J.~Schmidhuber, and F.~Cummins, ``Learning to {{Forget}}:
  {{Continual Prediction}} with {{LSTM}},'' {\em Neural Computation} {\bf 12},
  2451--2471  (2000).

\bibitem{borrelliPredictingTemporalDynamics2022}
G.~Borrelli, L.~Guastoni, H.~Eivazi, {\em et~al.}, ``Predicting the temporal
  dynamics of turbulent channels through deep learning,'' {\em International
  Journal of Heat and Fluid Flow} {\bf 96}, 109010  (2022).

\bibitem{zhengDisruptionPredictorBased2020}
W.~Zheng, Q.~Q. Wu, M.~Zhang, {\em et~al.}, ``Disruption predictor based on
  neural network and anomaly detection on {{J-TEXT}},'' {\em Plasma Physics and
  Controlled Fusion} {\bf 62}, 045012  (2020).

\bibitem{guoDisruptionPredictionEAST2021}
B.~H. Guo, D.~L. Chen, B.~Shen, {\em et~al.}, ``Disruption prediction on
  {{EAST}} tokamak using a deep learning algorithm,'' {\em Plasma Physics and
  Controlled Fusion} {\bf 63}, 115007  (2021).

\bibitem{liuHierarchicalDeepLearning2022}
Y.~Liu, J.~N. Kutz, and S.~L. Brunton, ``Hierarchical deep learning of
  multiscale differential equation time-steppers,'' {\em Philosophical
  Transactions of the Royal Society A: Mathematical, Physical and Engineering
  Sciences} {\bf 380}, 20210200  (2022).

\bibitem{jaegerHarnessingNonlinearityPredicting2004}
H.~Jaeger and H.~Haas, ``Harnessing {{Nonlinearity}}: {{Predicting Chaotic
  Systems}} and {{Saving Energy}} in {{Wireless Communication}},'' {\em
  Science} {\bf 304}, 78--80  (2004).

\bibitem{doanPhysicsinformedEchoState2020}
N.~A.~K. Doan, W.~Polifke, and L.~Magri, ``Physics-informed echo state
  networks,'' {\em Journal of Computational Science} {\bf 47}, 101237  (2020).

\bibitem{cuomoScientificMachineLearning2022}
S.~Cuomo, V.~S. Di~Cola, F.~Giampaolo, {\em et~al.}, ``Scientific {{Machine
  Learning Through Physics}}{\textendash}{{Informed Neural Networks}}:
  {{Where}} we are and {{What}}'s {{Next}},'' {\em Journal of Scientific
  Computing} {\bf 92}, 88  (2022).

\bibitem{wangNASPINNNeuralArchitecture2024}
Y.~Wang and L.~Zhong, ``{{NAS-PINN}}: {{Neural}} architecture search-guided
  physics-informed neural network for solving {{PDEs}},'' {\em Journal of
  Computational Physics} {\bf 496}, 112603  (2024).

\bibitem{luLearningNonlinearOperators2021}
L.~Lu, P.~Jin, G.~Pang, {\em et~al.}, ``Learning nonlinear operators via
  {{DeepONet}} based on the universal approximation theorem of operators,''
  {\em Nature Machine Intelligence} {\bf 3}, 218--229  (2021).

\bibitem{liNeuralOperatorGraph2020}
Z.~Li, N.~Kovachki, K.~Azizzadenesheli, {\em et~al.}, ``Neural {{Operator}}:
  {{Graph Kernel Network}} for {{Partial Differential Equations}},''  (2020).

\bibitem{liFourierNeuralOperator2021}
Z.~Li, N.~Kovachki, K.~Azizzadenesheli, {\em et~al.}, ``Fourier {{Neural
  Operator}} for {{Parametric Partial Differential Equations}},''  (2021).

\bibitem{caoLNOLaplaceNeural2023}
Q.~Cao, S.~Goswami, and G.~E. Karniadakis, ``{{LNO}}: {{Laplace Neural
  Operator}} for {{Solving Differential Equations}},''  (2023).

\bibitem{genevaTransformersModelingPhysical2022}
N.~Geneva and N.~Zabaras, ``Transformers for modeling physical systems,'' {\em
  Neural Networks} {\bf 146}, 272--289  (2022).

\bibitem{bruntonDiscoveringGoverningEquations2016}
S.~L. Brunton, J.~L. Proctor, and J.~N. Kutz, ``Discovering governing equations
  from data by sparse identification of nonlinear dynamical systems,'' {\em
  Proceedings of the National Academy of Sciences} {\bf 113}, 3932--3937
  (2016).

\bibitem{rudyDatadrivenDiscoveryPartial2017}
S.~H. Rudy, S.~L. Brunton, J.~L. Proctor, {\em et~al.}, ``Data-driven discovery
  of partial differential equations,'' {\em Science Advances} {\bf 3}, e1602614
   (2017).

\bibitem{alvesDatadrivenDiscoveryReduced2022}
E.~P. Alves and F.~Fiuza, ``Data-driven discovery of reduced plasma physics
  models from fully kinetic simulations,'' {\em Physical Review Research} {\bf
  4}, 033192  (2022).

\bibitem{kaptanogluPermanentMagnetOptimizationStellarators2022}
A.~A. Kaptanoglu, T.~Qian, F.~Wechsung, {\em et~al.}, ``Permanent-{{Magnet
  Optimization}} for {{Stellarators}} as {{Sparse Regression}},'' {\em Physical
  Review Applied} {\bf 18}, 044006  (2022).

\bibitem{kaptanogluSparseRegressionPlasma2023}
A.~A. Kaptanoglu, C.~Hansen, J.~D. Lore, {\em et~al.}, ``Sparse regression for
  plasma physics,'' {\em Physics of Plasmas} {\bf 30}, 033906  (2023).

\bibitem{kaiserSparseIdentificationNonlinear2018}
E.~Kaiser, J.~N. Kutz, and S.~L. Brunton, ``Sparse identification of nonlinear
  dynamics for model predictive control in the low-data limit,'' {\em
  Proceedings of the Royal Society A: Mathematical, Physical and Engineering
  Sciences} {\bf 474}, 20180335  (2018).

\bibitem{loreTimedependentSOLPSITERSimulations2023}
J.~D. Lore, S.~D. Pascuale, P.~Laiu, {\em et~al.}, ``Time-dependent
  {{SOLPS-ITER}} simulations of the tokamak plasma boundary for model
  predictive control using {{SINDy}}*,'' {\em Nuclear Fusion} {\bf 63}, 046015
  (2023).

\bibitem{settlesActiveLearningLiterature2009}
B.~Settles, ``Active {{Learning Literature Survey}},'' technical {{Report}},
  {University of Wisconsin-Madison Department of Computer Sciences}  (2009).

\bibitem{diawMultiscaleSimulationPlasma2020}
A.~Diaw, K.~Barros, J.~Haack, {\em et~al.}, ``Multiscale simulation of plasma
  flows using active learning,'' {\em Physical Review E} {\bf 102}(2), 023310
  (2020).

\bibitem{bozinovskiReminderFirstPaper2020}
S.~Bozinovski, ``Reminder of the {{First Paper}} on {{Transfer Learning}} in
  {{Neural Networks}}, 1976,'' {\em Informatica} {\bf 44}  (2020).

\bibitem{humbirdTransferLearningDriven2022}
K.~D. Humbird and J.~L. Peterson, ``Transfer learning driven design
  optimization for inertial confinement fusion,'' {\em Physics of Plasmas} {\bf
  29}, 102701  (2022).

\bibitem{gilpinExplainingExplanationsOverview2018}
L.~H. Gilpin, D.~Bau, B.~Z. Yuan, {\em et~al.}, ``Explaining {{Explanations}}:
  {{An Overview}} of {{Interpretability}} of {{Machine Learning}},'' in {\em
  2018 {{IEEE}} 5th {{International Conference}} on {{Data Science}} and
  {{Advanced Analytics}} ({{DSAA}})},  80--89  (2018).

\bibitem{ribeiroWhyShouldTrust2016}
M.~T. Ribeiro, S.~Singh, and C.~Guestrin, ``"{{Why Should I Trust You}}?":
  {{Explaining}} the {{Predictions}} of {{Any Classifier}},'' in {\em
  Proceedings of the 22nd {{ACM SIGKDD International Conference}} on
  {{Knowledge Discovery}} and {{Data Mining}}},  {\em {{KDD}} '16}, 1135--1144,
  {Association for Computing Machinery}, ({New York, NY, USA})  (2016).

\bibitem{wangMachineLearningExplainable2021}
C.-Y. Wang, T.-S. Ko, and C.-C. Hsu, ``Machine {{Learning}} with {{Explainable
  Artificial Intelligence Vision}} for {{Characterization}} of {{Solution
  Conductivity Using Optical Emission Spectroscopy}} of {{Plasma}} in {{Aqueous
  Solution}},'' {\em Plasma Processes and Polymers} {\bf 18}(12), 2100096
  (2021).

\bibitem{lundbergUnifiedApproachInterpreting2017}
S.~M. Lundberg and S.-I. Lee, ``A unified approach to interpreting model
  predictions,'' in {\em Proceedings of the 31st {{International Conference}}
  on {{Neural Information Processing Systems}}},  {\em {{NIPS}}'17},
  4768--4777, {Curran Associates Inc.}, ({Red Hook, NY, USA})  (2017).

\bibitem{wangRungeKuttaNeuralNetwork1998}
Y.-J. Wang and C.-T. Lin, ``Runge-{{Kutta}} neural network for identification
  of dynamical systems in high accuracy,'' {\em IEEE Transactions on Neural
  Networks} {\bf 9}, 294--307  (1998).

\bibitem{zhuConvolutionalNeuralNetworks2023}
M.~Zhu, B.~Chang, and C.~Fu, ``Convolutional {{Neural Networks}} combined with
  {{Runge-Kutta Methods}},'' {\em Neural Computing and Applications} {\bf 35},
  1629--1643  (2023).

\bibitem{depascualeDatadrivenLinearTime2022}
S.~De~Pascuale, D.~L. Green, and J.~D. Lore, ``Data-driven linear time advance
  operators for the acceleration of plasma physics simulation,'' {\em Physics
  of Plasmas} {\bf 29}, 113903  (2022).

\bibitem{kadupitiyaSolvingNewtonEquations2022}
J.~C.~S. Kadupitiya, G.~C. Fox, and V.~Jadhao, ``Solving {{Newton}}'s equations
  of motion with large timesteps using recurrent neural networks based
  operators,'' {\em Machine Learning: Science and Technology} {\bf 3}, 025002
  (2022).

\bibitem{sanchez-gonzalezGraphNetworksLearnable2018}
A.~{Sanchez-Gonzalez}, N.~Heess, J.~T. Springenberg, {\em et~al.}, ``Graph
  {{Networks}} as {{Learnable Physics Engines}} for {{Inference}} and
  {{Control}},'' in {\em Proceedings of the 35th {{International Conference}}
  on {{Machine Learning}}},  4470--4479, {PMLR}  (2018).

\bibitem{sanchez-gonzalezHamiltonianGraphNetworks2019}
A.~{Sanchez-Gonzalez}, V.~Bapst, K.~Cranmer, {\em et~al.}, ``Hamiltonian
  {{Graph Networks}} with {{ODE Integrators}},''  (2019).

\bibitem{liGraphNeuralNetworks2022}
Z.~Li, K.~Meidani, P.~Yadav, {\em et~al.}, ``Graph {{Neural Networks
  Accelerated Molecular Dynamics}},'' {\em arXiv:2112.03383 [physics]}
  (2022).

\bibitem{bruntonMachineLearningFluid2020}
S.~L. Brunton, B.~R. Noack, and P.~Koumoutsakos, ``Machine {{Learning}} for
  {{Fluid Mechanics}},'' {\em Annual Review of Fluid Mechanics} {\bf 52}(1),
  477--508  (2020).

\bibitem{vinuesaEnhancingComputationalFluid2022a}
R.~Vinuesa and S.~L. Brunton, ``Enhancing computational fluid dynamics with
  machine learning,'' {\em Nature Computational Science} {\bf 2}, 358--366
  (2022).

\bibitem{parkPredictiveControlPlasma2020}
S.~Park, Y.~Jang, T.~Cha, {\em et~al.}, ``Predictive control of the plasma
  processes in the {{OLED}} display mass production referring to the
  discontinuity qualifying {{PI-VM}},'' {\em Physics of Plasmas} {\bf 27},
  083507  (2020).

\bibitem{degraveMagneticControlTokamak2022}
J.~Degrave, F.~Felici, J.~Buchli, {\em et~al.}, ``Magnetic control of tokamak
  plasmas through deep reinforcement learning,'' {\em Nature} {\bf 602},
  414--419  (2022).

\bibitem{grassiReducingComplexityChemical2022}
T.~Grassi, F.~Nauman, J.~P. Ramsey, {\em et~al.}, ``Reducing the complexity of
  chemical networks via interpretable autoencoders,'' {\em Astronomy \&
  Astrophysics} {\bf 668}, A139  (2022).

\bibitem{tangReducedOrderModel2022}
K.~S. Tang and M.~Turk, ``Reduced {{Order Model}} for {{Chemical Kinetics}}:
  {{A}} case study with {{Primordial Chemical Network}},''  (2022).

\bibitem{chiavazzoReducedModelsChemical2014}
E.~Chiavazzo, C.~W. Gear, C.~J. Dsilva, {\em et~al.}, ``Reduced {{Models}} in
  {{Chemical Kinetics}} via {{Nonlinear Data-Mining}},'' {\em Processes} {\bf
  2}, 112--140  (2014).

\bibitem{goswamiLearningStiffChemical2023}
S.~Goswami, A.~D. Jagtap, H.~Babaee, {\em et~al.}, ``Learning stiff chemical
  kinetics using extended deep neural operators,''  (2023).

\bibitem{campoliAssessmentMachineLearning2022}
L.~Campoli, E.~Kustova, and P.~Maltseva, ``Assessment of {{Machine Learning
  Methods}} for {{State-to-State Approach}} in {{Nonequilibrium Flow
  Simulations}},'' {\em Mathematics} {\bf 10}, 928  (2022).

\bibitem{savareseMachineLearningClustering2023}
M.~Savarese, A.~Cuoci, W.~De~Paepe, {\em et~al.}, ``Machine learning clustering
  algorithms for the automatic generation of chemical reactor networks from
  {{CFD}} simulations,'' {\em Fuel} {\bf 343}, 127945  (2023).

\bibitem{maoDeepFlameDeepLearning2022}
R.~Mao, M.~Lin, Y.~Zhang, {\em et~al.}, ``{{DeepFlame}}: {{A}} deep learning
  empowered open-source platform for reacting flow simulations,''  (2022).

\bibitem{OpenFOAMFoundationLtd2023}
``The {{OpenFOAM Foundation Ltd}}, {{OpenFOAM}}: {{Open Source Field
  Operation}} and {{Manipulation}}.'' www.openfoam.org  (2023).

\bibitem{collobertTorch7MatlablikeEnvironment2011}
R.~Collobert, K.~Kavukcuoglu, and C.~Farabet, ``Torch7: {{A Matlab-like
  Environment}} for {{Machine Learning}},'' in {\em {{NIPS}} 2011 {{Workshop
  BigLearn}}},   (2011).

\bibitem{goodwinCanteraUserGuide2002}
D.~G. Goodwin, ``Cantera {{C}}++ {{User}}'s {{Guide}},'' tech. rep.,
  {California Institute of Technology}  (2002).

\bibitem{mouBridgingComplexityGap2023}
T.~Mou, H.~S. Pillai, S.~Wang, {\em et~al.}, ``Bridging the complexity gap in
  computational heterogeneous catalysis with machine learning,'' {\em Nature
  Catalysis} {\bf 6}, 122--136  (2023).

\bibitem{chenMachinelearningAtomicSimulation2023}
D.~Chen, C.~Shang, and Z.-P. Liu, ``Machine-learning atomic simulation for
  heterogeneous catalysis,'' {\em npj Computational Materials} {\bf 9}, 1--9
  (2023).

\bibitem{phelpsColdcathodeDischargesBreakdown1999}
A.~V. Phelps and Z.~L. Petrovic, ``Cold-cathode discharges and breakdown in
  argon: Surface and gas phase production of secondary electrons,'' {\em Plasma
  Sources Science and Technology} {\bf 8}(3), R21  (1999).

\bibitem{dakshaComputationallyAssistedSpectroscopic2016}
M.~Daksha, B.~Berger, E.~Schuengel, {\em et~al.}, ``A computationally assisted
  spectroscopic technique to measure secondary electron emission coefficients
  in radio frequency plasmas,'' {\em Journal of Physics D: Applied Physics}
  {\bf 49}(23), 234001  (2016).

\bibitem{schulzeComputationallyAssistedTechnique2022}
C.~Schulze, Z.~Donk{\'o}, and J.~Benedikt, ``A computationally assisted
  technique to measure material-specific surface coefficients in capacitively
  coupled plasmas based on characteristics of the ion flux-energy distribution
  function,'' {\em Plasma Sources Science and Technology} {\bf 31}, 105017
  (2022).

\bibitem{changCalculationSecondaryElectron2018}
H.-Y. Chang, A.~Alvarado, and J.~Marian, ``Calculation of secondary electron
  emission yields from low-energy electron deposition in tungsten surfaces,''
  {\em Applied Surface Science} {\bf 450}, 190--199  (2018).

\bibitem{suMechanismsAugerinducedChemistry2009}
J.~T. Su and W.~A. Goddard, ``Mechanisms of {{Auger-induced}} chemistry derived
  from wave packet dynamics,'' {\em Proceedings of the National Academy of
  Sciences} {\bf 106}, 1001--1005  (2009).

\bibitem{pamperinIoninducedSecondaryElectron2018}
M.~Pamperin, F.~X. Bronold, and H.~Fehske, ``Ion-induced secondary electron
  emission from metal surfaces,'' {\em Plasma Sources Science and Technology}
  {\bf 27}(8), 084003  (2018).

\bibitem{bronoldInvariantEmbeddingApproach2022}
F.~X. Bronold and H.~Fehske, ``Invariant embedding approach to secondary
  electron emission from metals,'' {\em Journal of Applied Physics} {\bf 131},
  113302  (2022).

\bibitem{zieglerStoppingRangeIons1985a}
J.~F. Ziegler and J.~P. Biersack, ``The {{Stopping}} and {{Range}} of {{Ions}}
  in {{Matter}},'' in {\em Treatise on {{Heavy-Ion Science}}},  93--129,
  {Springer}, {New York, USA}  (1985).

\bibitem{wangDeepLearningInteratomic2019}
H.~Wang, X.~Guo, L.~Zhang, {\em et~al.}, ``Deep learning inter-atomic potential
  model for accurate irradiation damage simulations,'' {\em Applied Physics
  Letters} {\bf 114}, 244101  (2019).

\bibitem{niuMachinelearningInteratomicPotential2023}
H.~Niu, J.~Zhao, H.~Li, {\em et~al.}, ``A machine-learning interatomic
  potential to understand primary radiation damage of silicon,'' {\em
  Computational Materials Science} {\bf 218}, 111970  (2023).

\bibitem{caroGrowthMechanismOrigin2018}
M.~A. Caro, V.~L. Deringer, J.~Koskinen, {\em et~al.}, ``Growth {{Mechanism}}
  and {{Origin}} of {{High}} sp3 {{Content}} in {{Tetrahedral Amorphous
  Carbon}},'' {\em Physical Review Letters} {\bf 120}, 166101  (2018).

\bibitem{caroMachineLearningDriven2020}
M.~A. Caro, G.~Cs{\'a}nyi, T.~Laurila, {\em et~al.}, ``Machine learning driven
  simulated deposition of carbon films: {{From}} low-density to diamondlike
  amorphous carbon,'' {\em Physical Review B} {\bf 102}, 174201  (2020).

\bibitem{thompsonLAMMPSFlexibleSimulation2022}
A.~P. Thompson, H.~M. Aktulga, R.~Berger, {\em et~al.}, ``{{LAMMPS}} - a
  flexible simulation tool for particle-based materials modeling at the atomic,
  meso, and continuum scales,'' {\em Computer Physics Communications} {\bf
  271}, 108171  (2022).

\bibitem{limEvolutionMetastableStructures2020}
J.~S. Lim, J.~Vandermause, M.~A. {van Spronsen}, {\em et~al.}, ``Evolution of
  {{Metastable Structures}} at {{Bimetallic Surfaces}} from {{Microscopy}} and
  {{Machine-Learning Molecular Dynamics}},'' {\em Journal of the American
  Chemical Society} {\bf 142}, 15907--15916  (2020).

\bibitem{henkelmanDimerMethodFinding1999}
G.~Henkelman and H.~J{\'o}nsson, ``A dimer method for finding saddle points on
  high dimensional potential surfaces using only first derivatives,'' {\em The
  Journal of Chemical Physics} {\bf 111}, 7010--7022  (1999).

\bibitem{henkelmanClimbingImageNudged2000}
G.~Henkelman, B.~P. Uberuaga, and H.~J{\'o}nsson, ``A climbing image nudged
  elastic band method for finding saddle points and minimum energy paths,''
  {\em The Journal of Chemical Physics} {\bf 113}, 9901--9904  (2000).

\bibitem{henkelmanImprovedTangentEstimate2000}
G.~Henkelman and H.~J{\'o}nsson, ``Improved tangent estimate in the nudged
  elastic band method for finding minimum energy paths and saddle points,''
  {\em The Journal of Chemical Physics} {\bf 113}, 9978--9985  (2000).

\bibitem{frankePlasmaMDSMetadataSchema2020}
S.~Franke, L.~Paulet, J.~Sch{\"a}fer, {\em et~al.}, ``Plasma-{{MDS}}, a
  metadata schema for plasma science with examples from plasma technology,''
  {\em Scientific Data} {\bf 7}, 439  (2020).

\bibitem{carboneDataNeedsModeling2021}
E.~Carbone, W.~Graef, G.~Hagelaar, {\em et~al.}, ``Data {{Needs}} for
  {{Modeling Low-Temperature Non-Equilibrium Plasmas}}: {{The LXCat Project}},
  {{History}}, {{Perspectives}} and a {{Tutorial}},'' {\em Atoms} {\bf 9}, 16
  (2021).

\bibitem{siffaAdamantJSONSchemabased2022a}
I.~C. Siffa, J.~Sch{\"a}fer, and M.~M. Becker, ``Adamant: A {{JSON}}
  schema-based metadata editor for research data management workflows,'' {\em
  F1000Research} {\bf 11}, 475  (2022).

\bibitem{wilkinsonFAIRGuidingPrinciples2016}
M.~D. Wilkinson, M.~Dumontier, I.~J. Aalbersberg, {\em et~al.}, ``The {{FAIR
  Guiding Principles}} for scientific data management and stewardship,'' {\em
  Scientific Data} {\bf 3}, 160018  (2016).

\end{thebibliography}
\bibliographystyle{spiejour}   







\end{spacing}
\end{document}